\documentclass[twocolumn]{aastex61}

\newcommand{\lsim}{\raisebox{-0.13cm}{~\shortstack{$<$ \\[-0.07cm] $\sim$}}~}
\newcommand{\gsim}{\raisebox{-0.13cm}{~\shortstack{$>$ \\[-0.07cm] $\sim$}}~}

\received{...}
\revised{...}
\accepted{...}
\submitjournal{ApJ}

\shorttitle{The Evolution of Dusty and Non-Dusty Galaxies with Stellar Mass at $\lowercase{z}=2-6$ from SMUVS}
\shortauthors{Deshmukh et al.}
\begin{document}

\title{The Spitzer Matching Survey of the UltraVISTA Ultra-deep Stripes (SMUVS): the Evolution of Dusty and Non-Dusty Galaxies with Stellar Mass at $\lowercase{z}=2-6$}

\correspondingauthor{S. Deshmukh}
\email{deshmukh@astro.rug.nl}

\author{S. Deshmukh}
\affil{Kapteyn Astronomical Institute, University of Groningen, P.O. Box 800, 9700AV Groningen, The Netherlands}

\author{K. I. Caputi}
\affil{Kapteyn Astronomical Institute, University of Groningen, P.O. Box 800, 9700AV Groningen, The Netherlands}
\affil{Cosmic Dawn Center (DAWN), Niels Bohr Institute, University of Copenhagen, Juliane Maries vej 30, DK-2100 Copenhagen, Denmark}

\author{M. L. N. Ashby}
\affil{Harvard-Smithsonian Center for Astrophysics, 60 Garden St., Cambridge, MA 02138, USA}

\author{W.I. Cowley}
\affil{Kapteyn Astronomical Institute, University of Groningen, P.O. Box 800, 9700AV Groningen, The Netherlands}

\author{H. J. McCracken}
\affil{Institut d'Astrophysique de Paris, CNS \& UPMC, UMR 7095, 98 bis Boulevard Arago, F-75014, Paris, France}

\author{J. P. U. Fynbo}
\affil{Cosmic Dawn Center (DAWN), Niels Bohr Institute, University of Copenhagen, Juliane Maries vej 30, DK-2100 Copenhagen, Denmark}

\author{O. Le F\`{e}vre}
\affil{Aix Marseille Universit\'{e}, CNRS, LAM (Laboratoire d'Astrophysique de Marseille), UMR 7326, 13388, Marseille, France}

%\author{R. S. Somerville}
%\affil{Department of Physics and Astronomy, Rutgers University, The State University of New Jersey, 136 Frelinghuysen Road, Piscataway, NJ 08854, USA}

\author{B. Milvang-Jensen}
\affil{Dark Cosmology Centre, Niels Bohr Institute, University of Copenhagen, Juliane Maries Vej 30, DK-2100 Copenhagen, Denmark}

\author{O. Ilbert}
\affil{Aix Marseille Universit\'{e}, CNRS, LAM (Laboratoire d'Astrophysique de Marseille), UMR 7326, 13388, Marseille, France}

\begin{abstract}

The Spitzer Matching Survey of the UltraVISTA Ultra-deep Stripes (SMUVS) has obtained the largest ultra-deep {\em Spitzer} maps to date in a single field of the sky. We considered the sample of about 66,000 SMUVS sources at $z=2-6$ to investigate the evolution of dusty and non-dusty galaxies with stellar mass through the analysis of the galaxy stellar mass function (GSMF), extending previous analyses about one decade in stellar mass and up to $z=6$. We further divide our non-dusty galaxy sample with rest-frame optical colours to isolate red quiescent (`passive') galaxies. At each redshift, we identify a characteristic stellar mass in the GSMF above which dusty galaxies dominate, or are at least as important as non-dusty galaxies.  Below that stellar mass, non-dusty galaxies comprise about 80\% of all sources, at all redshifts except at $z=4-5$.  The percentage of dusty galaxies at $z=4-5$ is unusually high: 30-40\% for $M_{*}=10^9 - 10^{10.5} \, \rm M_\odot$ and  $>80\%$ at $M_*>10^{11} \, \rm M_\odot$, which indicates that dust obscuration is of major importance in this cosmic period. The overall percentage of massive ($\log_{10} (M_*/M_\odot)>10.6$) galaxies that are quiescent increases with decreasing redshift, reaching $>30\%$ at $z\sim2$. Instead, the quiescent percentage among intermediate-mass  galaxies (with $\log_{10} (M_*/M_\odot)=9.7-10.6$) stays roughly constant at a $\sim 10\%$ level. Our results indicate that massive and intermediate-mass galaxies clearly have different evolutionary paths in the young Universe, and are consistent with the scenario of galaxy downsizing.

\end{abstract}




\section{Introduction} 
\label{sec:intro}

Dust has increasingly been recognised as a key element in galaxy growth, being the by-product of stellar evolution and a catalyst for the formation of molecular hydrogen \citep[e.g.,][]{1963ApJ...138..393G, 1971ApJ...163..155H,2012MNRAS.424.2961G}. Because of this, analyzing the presence of dust in galaxies at different redshifts can shed light on fundamental aspects of galaxy evolution \citep[e.g.,][]{cal17,2017MNRAS.471.3152P}. Although the dust content in present-day galaxies is known to be moderate, the role of dust was much more important in the past, as has tentatively been inferred already two decades ago \citep[e.g.,][]{hug98,ade00}, and more recently shown by multiple observational studies conducted with mid and far-infrared (IR) telescopes \citep[e.g.,][]{cap07,gru10,mag11}. 

Despite significant progress, the nature of dust and the presence of dusty galaxies at high redshifts is not well understood. While theoretical models are successful in producing massive dusty/passive galaxies, they find it challenging to reproduce their number counts, physical properties and their $z=0$ stellar mass functions \textit{simultaneously}. (e.g., \citealp{2011MNRAS.417.2676G, 2014PhR...541...45C,2017MNRAS.470.1050F}; see also Sec 4.1.3 in \citealp{2015ARA&A..53...51S}).

The task of determining the dust content of galaxies at moderate-high redshifts has proven to be quite challenging. Given the limited sensitivity of mid-/far-IR telescopes, identifying the presence of dust in a wide range of galaxies at different redshifts requires a different approach. An alternative option to direct observations is to model the galaxy spectral energy distribution (SED), as dust extinction is usually considered as a free parameter for the fitting. This method has been followed by multiple authors from low to high redshifts \citep[e.g.,][]{pan09,cuc12}.  In parallel, either through SED fitting or colour selections, different works have attempted to identify quiescent galaxies, i.e. galaxies whose levels of star formation activity can be considered negligible with respect to the amount of stars formed at previous times \citep[e.g.,][]{kaj11,cas13,som14,str14,mar16}. 
.

The {\em Spitzer Matching Survey of the Ultra-VISTA ultra-deep Stripes} (SMUVS; M.~Ashby et al., 2018, in preparation) is a {\em Spitzer} \citep{wer04}  Exploration Science Program,  which has obtained ultra-deep $3.6$ and $4.5 \, \rm \mu m$ imaging with the Infrared Array Camera \citep[IRAC;][]{faz04} in the COSMOS field \citep{sco07}. SMUVS has been designed to complement the UltraVISTA ultra-deep near-IR survey \citep{mcc12} in the region with deepest optical coverage. Until today, SMUVS is the largest quasi-contiguous {\em Spitzer} field suitable to study the high-$z$ Universe. Its unique combination of area and depth allows us to investigate different aspects of galaxy evolution with an unprecedented level of statistics and dynamic range at high redshifts.

In this paper we analyze the large SMUVS galaxy sample containing a total of $\sim$66,000 galaxies at $z=2-6$. We conduct an unprecedented analysis of the evolution of galaxies with and without significant dust extinction (dusty and and non-dusty galaxies hereafter) as a function of stellar mass spanning the period between $\sim 1$ and 3.2~Gyr after the Big Bang. In parallel, in other companion papers we analyse the clustering properties of the SMUVS galaxies over a similar redshift range \citep{cow18}, and study star formation in galaxies at $z=4-5$ as inferred from their $\rm H\alpha$ excess in the IRAC $3.6 \, \rm \mu m$ band \citep{cap17}. Besides, an independent work included the SMUVS data to trace the progenitors of present-day massive galaxies out to $z=5$ \citep{2017ApJ...837..147H}.

This paper is organised as follows. In Section~\ref{sec:data} we describe the utilised datasets and source catalogue construction and in Section~\ref{sec:zphotstm} we explain our derivation of galaxy properties, including photometric redshifts and stellar masses. We present the GSMF of dusty and non-dusty galaxies in Section~\ref{sec:gsmf}, and analyse the overall number densities of dusty/non-dusty sources, as well as the evolution to quiescence, among massive and intermediate-mass galaxies in Section~\ref{sec:evolquies}. Finally, in Section~\ref{sec:conclus} we present our concluding remarks. Throughout this paper we adopt a cosmology  with $\rm H_0=70 \,{\rm km \, s^{-1} Mpc^{-1}}$, $\rm \Omega_M=0.3$ and $\rm \Omega_\Lambda=0.7$. All magnitudes and fluxes are total, with magnitudes referring to the AB system \citep{oke83}. Stellar masses correspond to a \citet{cha03} initial mass function (IMF).

\section{Datasets and Source Catalogue} \label{sec:data}

As part of the SMUVS program (PI Caputi; M.~Ashby et al., 2018), we have collected ultra-deep {\em Spitzer} 3.6 and 4.5~$\rm \mu m$ data in the COSMOS field \citep{sco07}, over an area overlapping the three UltraVISTA ultra-deep stripes \citep{mcc12} with the deepest optical coverage from the Subaru telescope \citep{tan07}. The SMUVS mosaics considered in this paper correspond to the almost final depth of the survey, which reaches on average an integration time of $\sim 25 \, \rm h$/pointing \citep[including IRAC ancillary data in COSMOS;][]{san07,ash13,ste14,ash15}. These long integration times, coupled to the large IRAC point-spread-function (PSF) full width at half maximum (FWHM), which is about 1.9~arcsec, imply that the resulting SMUVS images suffer from severe source confusion. Therefore, we apply a technique that includes source deblending in order to measure the IRAC photometry. We proceed as follows.

First, we construct UltraVISTA $HK_s$ average stack maps of the three relevant ultra-deep stripes, which we use as priors in the IRAC PSF-fitting. The UltraVISTA data considered here correspond to the third data release (DR3), which in the ultra-deep stripes reaches an average depth of $K_s=24.9 \pm 0.1$ and $H=25.1\pm 0.1$  ($2^{\prime\prime}$ diameter; $5\sigma$)\footnote{see http://www.eso.org/sci/observing/phase3/data\_releases/ \\ uvista\_dr3.pdf}. We extract sources the $HK_s$ sources using the software SExtractor \citep{ber96} with a detection threshold of 1.5$\sigma$ over 5 contiguous pixels.  Using these source positions, we measure their photometry on the SMUVS 3.6 and 4.5~$\rm \mu m$ mosaics, applying a point-spread-function (PSF)-fitting technique with the DAOPHOT package on IRAF. This PSF-fitting technique (applied in the task ``allstar'') consists of fitting the photometry of groups of sources simultaneously and iteratively until the fluxes are deblended and the residuals are minimized. In order to maximize the number of detected sources through PSF fitting, we run ``allstar'' twice: a first run is done on each original image, and a second pass is done on each residual image.

For the PSF-fitting technique, we make use of empirical images of the PSF, which  we construct from stars in the field, in each stripe separately. With the PSF-fitting algorithm, we achieve convergence for $\sim70\%$ of the sources, after the two passes described above. This degree of convergence is normal, given that we are trying to PSF-fit sources down to very faint levels (based on the known UltraVISTA coordinates). For the remaining  $\sim30\%$ sources,  we directly measure IRAC aperture fluxes in $2.4 \, \rm arcsec$-diameter circular apertures at the UltraVISTA positions.  We correct these aperture fluxes to total fluxes by multiplying them by a factor of 2.13, which is determined from the curves of growth of stars in the field. In total, we find that 95-96\% of all UltraVISTA ultra-deep sources are detected in at least one IRAC band, and 93-94\% in both bands. In the following, we refer to the UltraVISTA ultra-deep sources with at least one IRAC detection as the `SMUVS sources'.

As we explain in detail in Section~\ref{sec:zphotstm}, we do not use the IRAC photometry for the SED fitting of sources with potentially significant light contamination in any of the IRAC bands. This applies to $<14\%$ of our sources. This criterion allows us to minimize the impact that any IRAC light contamination can have on the derived source properties.

\begin{figure*}[ht!]
\center{
\includegraphics[width=1\linewidth, keepaspectratio]{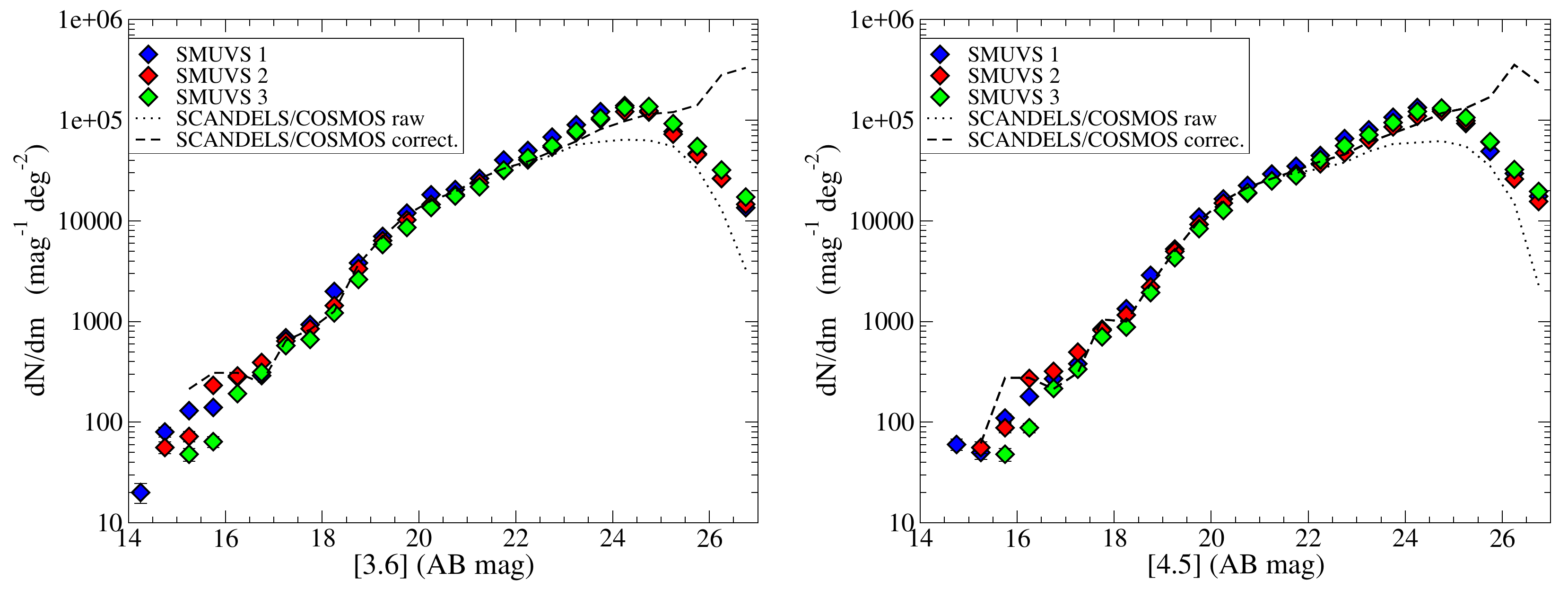}
\caption{SMUVS number counts obtained using the UltraVISTA $HK_s$ source positions as priors. Left: IRAC channel 1 ($3.6 \, \rm \mu m$).   Right: IRAC channel 2 ($4.5 \, \rm \mu m$). The corrected SCANDELS/COSMOS number counts are based on the \cite{2012ApJ...752..113H} models. } \label{fig_numc}
}
\end{figure*}

Fig.~\ref{fig_numc} shows the SMUVS  3.6 and 4.5~$\rm \mu m$ number counts obtained using the UltraVISTA sources as priors. The results for the three stripes are in very good agreement among themselves, and indicate that our resulting SMUVS catalogues are 80\%(50\%) complete at [3.6] and [4.5]=25.5 (26.0) total magnitudes. Note that this completeness is higher than that obtained with the raw counts in the SCANDELS/COSMOS field \citep{ash15}, in spite of the latter images being deeper on average. This is because the SCANDELS/COSMOS number counts have been obtained from a direct source extraction using no priors, and because both a detection at 3.6 and $4.5 \, \rm \mu m$ have been imposed to consider a source reliable. Our prior UltraVISTA detection makes the criterion of two IRAC detections unnecessary to guarantee the source reliability.
 
\begin{figure}
 	\plotone{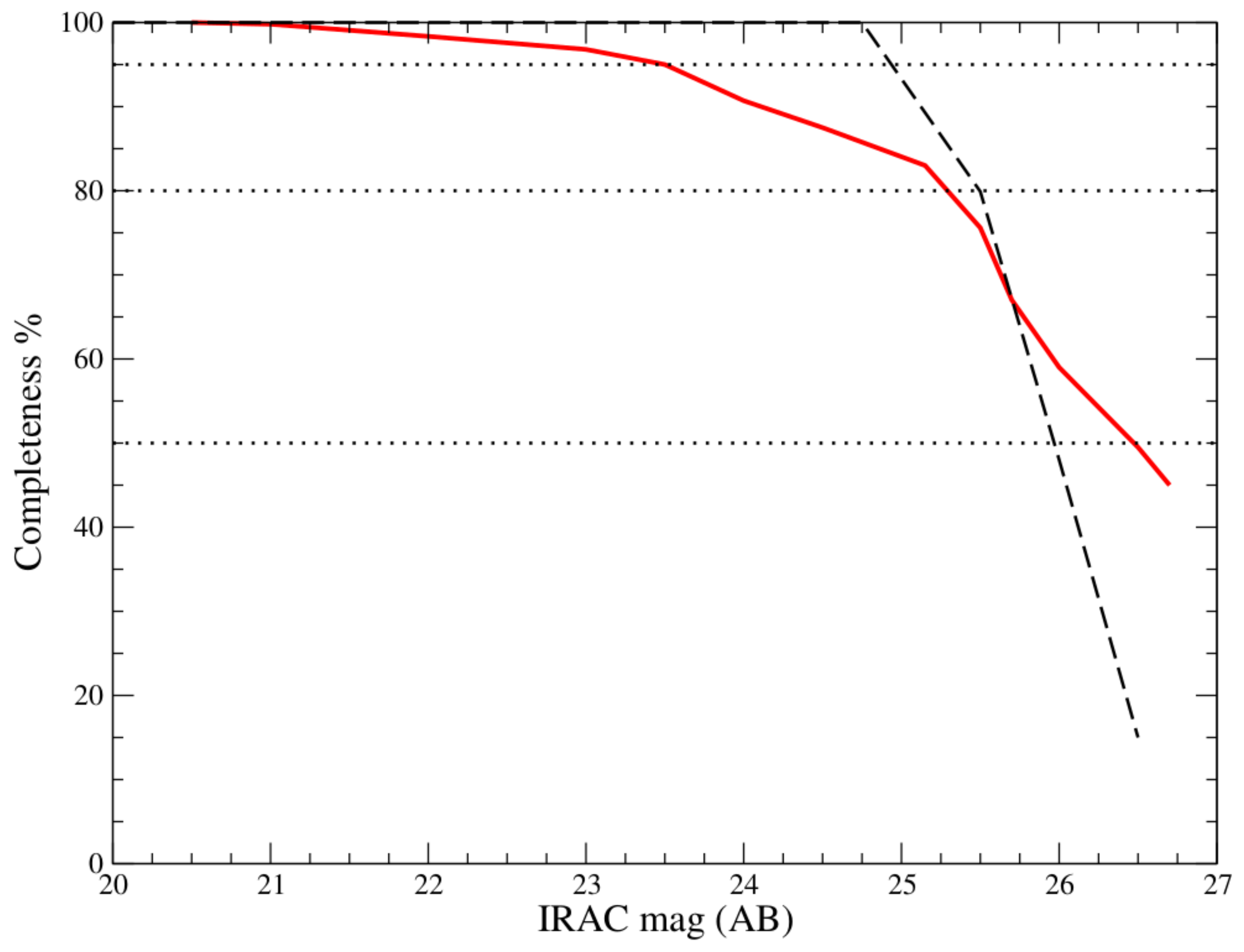}
 	\caption{SMUVS source completeness as derived from mock galaxy simulations (red solid line; see text for details). The completeness derived from the IRAC number count comparison with the \cite{2012ApJ...752..113H} models is also shown (black dashed line). The horizontal dotted lines indicate the 95\%, 80\% and 50\% completeness limits for reference.}
 	\label{fig: completeness}
 \end{figure}

We have independently checked the number counts completeness limits by performing simulations similarly to those in \cite{cap11}. However, in contrast to this work in which there was only a direct source extraction on the IRAC images, here we need to emulate our IRAC photometric extraction based on the UltraVISTA source priors. For this, in our simulations we proceeded as follows: we created a catalogue of 50,000 mock sources using the IRAF task `galllist', following a power-law distribution between magnitudes 17 and 28. We then created a set of 10 mock UltraVISTA HKs images (using IRAF \textsc{mkobjects}), based on the original HKs mosaics, in each of which we inserted 5,000 of the mock sources without repetition. We then ran SExtractor with the same parameter values used for the original HKs mosaics, and compiled the recovered mock sources. These recovered mock sources were used as priors to be inserted in the IRAC images. We created a set of mock IRAC mosaics, based on the original mosaics, in which we inserted no more than 500 of the recovered mock sources at a time (we did not add more sources per image to avoid altering the confusion properties). We then used the IRAF DAOPHOT package at the position of the known prior sources, as in our original methodology. To correspond the HKs magnitudes with the IRAC magnitudes, we have considered the colour distribution of the real SMUVS sources in different magnitude bins. %we considered that the median (HKs-IRAC) colours of our real sources are $\sim0.25-0.28$ mag. 
The final number count completeness levels as a function of mag are the product of the completeness obtained in the SExtracted recoverery of mock sources in the HKs-band mosaic, %and the completeness on the IRAC photometry measurements using priors. 
and the completeness on the IRAC photometry measurements using priors (where, at each IRAC magnitude, we applied weights to take into account the real source (HKs-IRAC) colour distributions). From this combined calculation, we get that at IRAC mag=25.5 (26.0) we have 76\% (59\%) completeness, %74\% (52\%) completeness,
 which is very similar to the completeness levels derived from comparison with the \cite{2012ApJ...752..113H} models (Figure \ref{fig: completeness}) .
 
For all the SMUVS sources, we measure 2-arcsec diameter circular photometry on 26 broad, intermediate and narrow bands, namely CFHT $U$ band; Subaru $B$, $V$, $r$, $i^+$, $z^+$, $z^{++}$, $IA427$, $IA464$, $IA484$, $IA505$, $IA527$, $IA574$, $IA624$, $IA679$, $IA709$, $IA738$, $IA767$, $IA827$, $NB711$ and $NB816$; {\em HST} $F814W$; and UltraVISTA $Y$, $J$, $H$ and $K_s$. We use SExtractor in dual-image mode with the UltraVISTA $HK_s$ stacks as detection images. We correct the measured aperture fluxes to total fluxes by applying point-source aperture corrections in each band. In addition, we correct all our photometry for Galactic extinction. To determine errors on the photometry, we perform empty-aperture statistics in different parts of each stripe.

\begin{figure}[ht!]
\center{
\includegraphics[width=0.7\linewidth, angle=270, keepaspectratio]{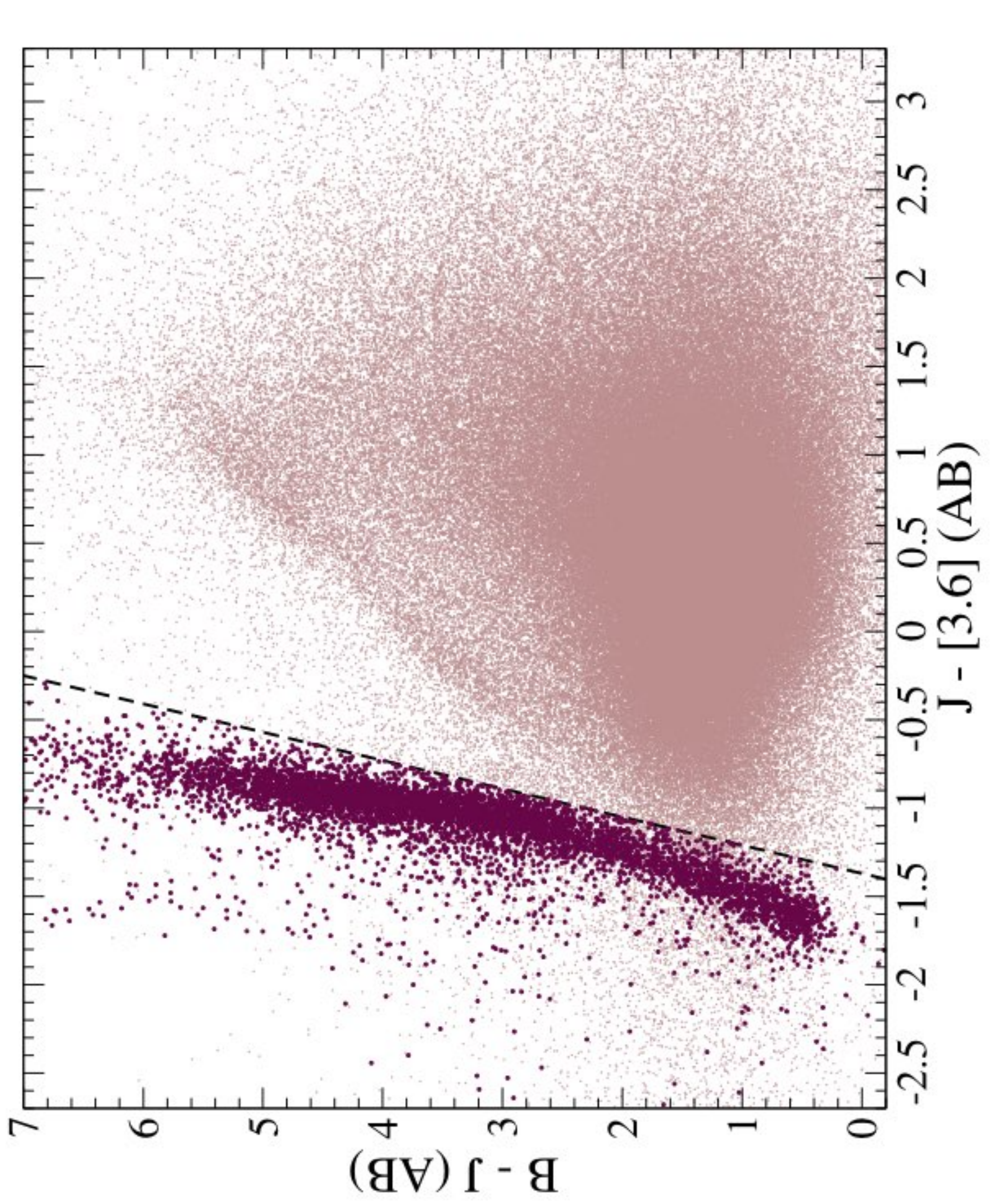}
\caption{$J$-[3.6]) vs. ($B$-$J$) colour-colour diagram for the SMUVS sources extracted with UltraVISTA $HK_s$ priors. Galactic stars appear clearly segregated on the left-hand side of this plot.} \label{fig_colst}
}
\end{figure}

Following \citet{cap11}, we clean our catalogue for galactic stars, using SExtractor's stellarity parameter and a ($J$-[3.6]) vs. ($B$-$J$) colour-colour diagnostic (Fig.~\ref{fig_colst}). We discard all sources with an $HK_s$-based stellarity parameter greater than 0.8 that lie on the stellar sequence clearly defined on the ($J$-[3.6]) versus ($B$-$J$) diagram. These rejected stars constitute $\sim 2\%$ of the original SMUVS sources detected using the UltraVISTA priors. We also mask regions of contaminated light around the brightest sources to obtain a clean catalogue of UltraVISTA ultra-deep sources with at least one IRAC-band detection, over a net area of 0.66~deg$^2$. This is the catalogue with 28-band photometry ($U$ through $K_s$ + IRAC) that we consider as input for the SED fitting analysis.

The PSF-fitting technique assumes that all sources are point-like. This is a reasonable assumption for virtually all IRAC sources with $[3.6]>21$~mag \citep[see Fig.~25 in][]{ash13}. Besides, this approach is consistent with all our other multi-wavelength photometry, measured on circular apertures (and corrected to total), which also implicitly assumes that all sources are point-like. Some other methods to derive IRAC photometry \citep[e.g.,][]{mer15} do take into account the shape of resolved sources, and this may be preferable for the study of low-$z$ galaxies, as many of them are resolved even in the IRAC bands. Here, however, we only focus on the analysis of $z>2$ sources, and thus the point-like source assumption can safely be adopted.

\section{SMUVS galaxy properties determined from SED fitting} \label{sec:zphotstm}

\subsection{Photometric Redshifts}

To perform the source SED fitting, we run the $\chi^2$-minimization code \textsc{LePhare} \citep{arn99,ilb06} on our catalogue with total fluxes, based on 2-arcsec aperture photometry for all bands $U$ through $K_s$, and obtained as described above for the IRAC bands. As in \citet{cap15}, in the case of non-detections we adopt $3\sigma$ flux upper limits in the broad bands, also determined from empty-aperture statistics (up to the $K_s$ band), and ignored narrow and intermediate bands, as well as any IRAC band with a non-detection. Note that this non-detection treatment is only done for true SExtractor non-detections. Each time that SExtractor extracts a meaningful (positive) flux, we leave that flux measurement, even if it has a large associated error bar. Within \textsc{LePhare}, we choose the following option: all SED templates that produce fluxes higher than the $3\sigma$ upper limits in the bands with non-detections are automatically discarded.

To minimize the chances of affecting the SED fitting due to contamination in the IRAC photometry, we impose the following: for the sources with an IRAC (3.6 or $4.5 \, \rm \mu m$) magnitude $>22$ having an IRAC neighbour with a magnitude $<23$ within less than 3~arscec radius,  we do not utilise the IRAC photometry in the SED fitting (we only used the 26 bands from $U$ through $K_s$). This situation applies to $<14$\% of all SMUVS sources, and only 12\% at $2<z<6$. Comparison of the photometric redshifts obtained with and without IRAC photometry for these sources, with respect to spectroscopic redshifts in COSMOS, indicate that excluding the IRAC photometry is the right approach in this case.

We use a series of synthetic templates from the \citet{bc03} library, namely a simple stellar population and different exponentially declining star formation histories (SFHs) with star-formation timescales $\tau=0.01, 0.1, 0.3, 1.0, 3.0, 5.0, 10.0$ and $15 \, \rm Gyr$.   Each synthetic spectrum is attenuated with the \citet{cal00} reddening law, leaving the colour excess as a free parameter with possible values $\rm E(B-V) = 0.0-1.0$ in steps 0.1. Adopting a finer colour excess grid does not have any significant impact on our results\footnote{For the low stellar-mass dusty galaxies, the differences in the GSMF with our adopted and a more refined extinction grids are larger than the error bars, but still very small ($\leq0.1-0.2$ dex).}. The Calzetti et al. reddening law appears to be the most suitable for high-z galaxies \citep[][]{2017arXiv171201292C} . We run \textsc{LePhare} with emission lines and iterate to obtain photometric zero-point corrections, which significantly improves the overall quality of our photometric redshifts, as determined from the comparison with spectroscopic redshifts, when available (see discussion below). All zero-point corrections are $\lsim 0.1$~mag in absolute value, except in the $V$-band for which we derive a correction of -0.18~mag.

\begin{figure}
\plotone{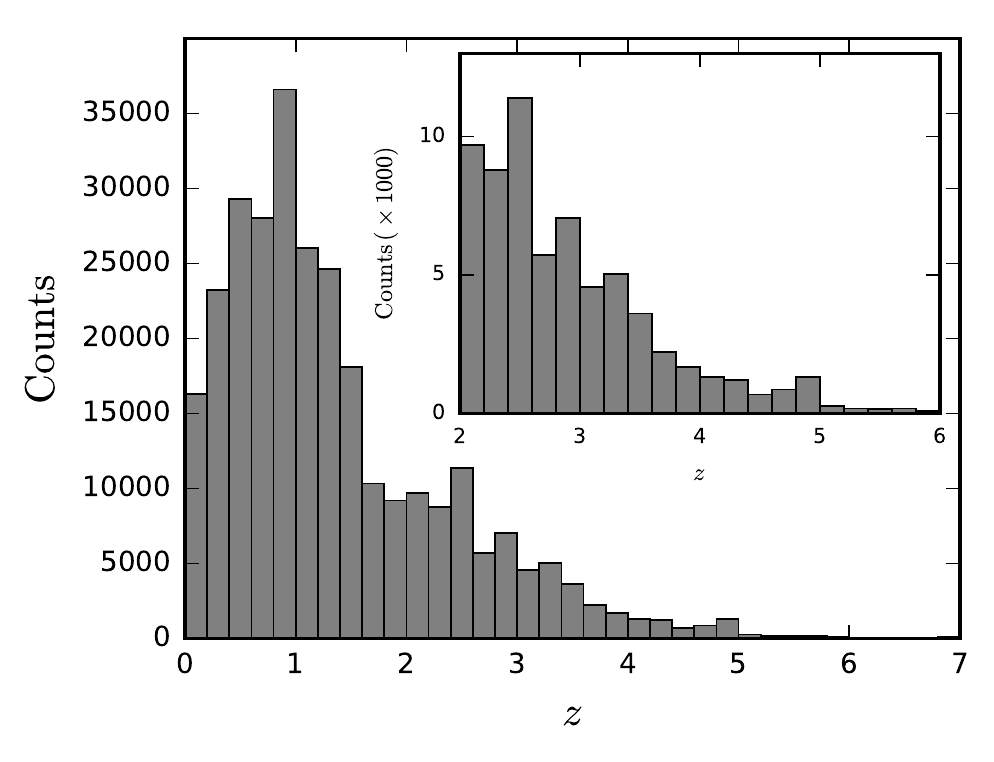}
\caption{Redshift distribution of our SMUVS galaxy sample. The inset zooms in the redshift range considered for scientific analysis in this paper.}
\label{fig:zphotdist}
\end{figure}

\begin{figure}
	\centering
	\plotone{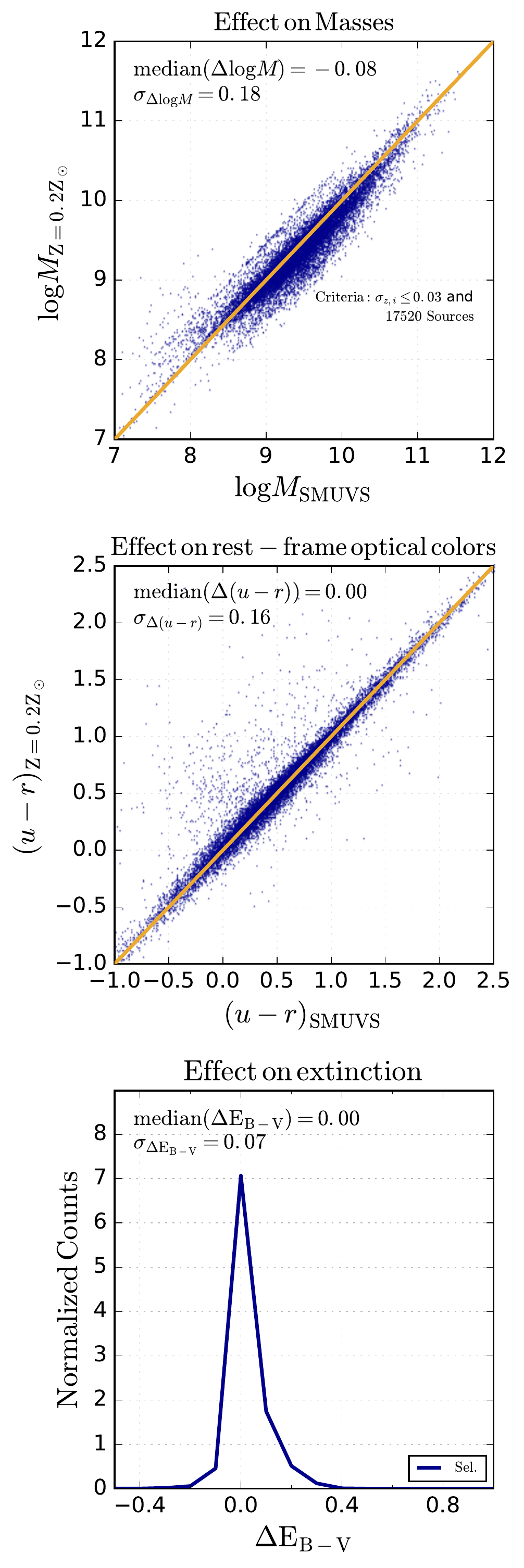}
	\caption{Impact of allowing sub-solar metallicities on masses, rest-frame optical colors and extinction. The comparison has been made for sources with $|z_{Z_\odot} - z_{0.2Z_\odot} | / (1+z_{Z_\odot}) \leq 0.03$, ensuring that we do not carry differences produced by very different redshifts.}
	\label{fig:zsun vs 02zsun}
\end{figure}

Our strategy for photometric redshift determination is as follows. We first run \textsc{LePhare} using only spectral templates with solar metallicity.  For galaxies with a primary redshift solution $z_{\rm phot}<5$, we consider that the best solar-metallicity template was the final best-fit model. Instead, for galaxies with a best-fit redshift $z\geq5$,  we re-ran \textsc{LePhare} using an equivalent set of templates with a sub-solar metallicity, namely $Z = 0.2\, \rm Z_\odot$. We compare the minimum reduced $\chi^2$ obtained with both metallicity runs, and adopt as best-fit model that providing the absolute smallest  $\chi^2$  value. In total, about 34\% of the $z_{\rm phot}\geq5$ prefer a model with sub-solar metallicity.  The redshift cut to consider only $Z = \rm Z_\odot$, or both $Z = \rm Z_\odot$ and $Z = 0.2\, \rm Z_\odot$, has been calibrated through the comparison of the resulting best photometric redshifts with spectroscopic redshifts, as described in the next paragraph. The approach of considering the possibility of sub-solar metallicities only at $z\geq5$ does not introduce any significant bias in our results. Figure \ref{fig:zsun vs 02zsun} shows the compared rest (u-r), E(B-V) and stellar mass values obtained when considering two possible metallicity values (0.2$Z_\odot$ and $Z_\odot$) also at $z<5$, versus our values obtained with fixed metallicity. These plots show that the biases in these properties are negligible or very small compared with the corresponding error bars, and the scatter is also small.

\begin{figure}
\plotone{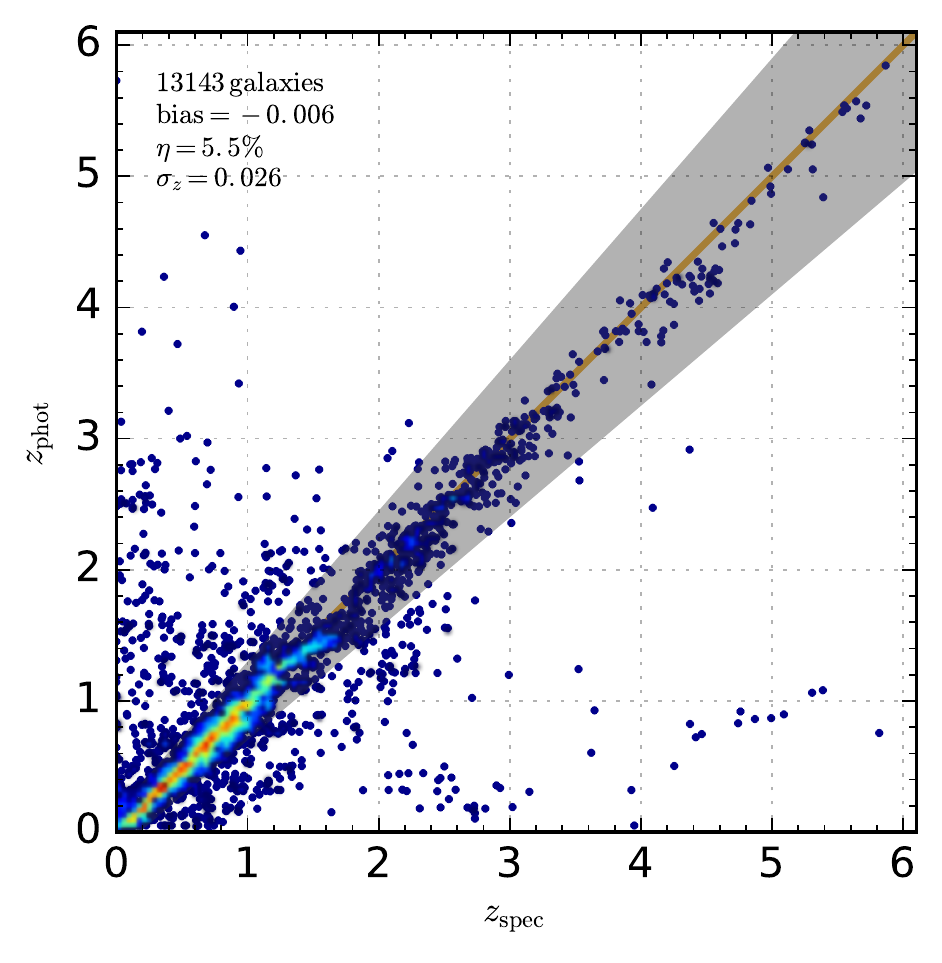}
\caption{Comparison between our $z_\mathrm{phot}$ and existing $z_\mathrm{spec}$ for $\sim 13,000$ SMUVS galaxies. The colour scale indicates the log-density of sources.}
\label{fig:zspecdiag}
\end{figure}

From \textsc{LePhare}'s runs we obtain photometric redshifts and stellar mass estimates for $> 99\%$ of our sources.  For the remaining $<1\%$, \textsc{LePhare} indicated that a stellar template yielded a lower minimum reduced $\chi^2$ than any galaxy template in the SED fitting. We discard these sources from our sample. We also exclude a small percentage of sources $<1\%$  because their best $z_{\rm phot}$ were incompatible with their detection at short wavelengths, i.e., they have either a $> 2\sigma$ $U$-band detection and a redshift $z_{\rm phot}>3.6$; or a $> 2\sigma$ $B_j$-band detection and $z_{\rm phot}>4.6$; or  a $> 2\sigma$ $V_j$-band detection and $z_{\rm phot}>5.6$  \citep[see][]{cap15}. Our final SMUVS output catalogue with photometric redshift determinations contains 288,003 galaxies. In Fig.~\ref{fig:zphotdist} we show the resulting redshift distribution. The inset zooms in the redshift range considered for scientific analysis in this paper, i.e., $z=2-6$, which contains about 66,000 SMUVS galaxies.
  
We use the large amount of spectroscopic data available in COSMOS  \citep[e.g.,][]{lil07,com15,lef15} to assess the quality of our obtained photometric redshifts. Fig.~\ref{fig:zspecdiag} shows the resulting $z_\mathrm{phot}$ vs. $z_\mathrm{spec}$ diagnostic, which is based on more than 13,000 galaxies (including 627 at $z_{\rm spec}>2$). These make for $\sim 4\%$  of our SMUVS sources ($\sim1\%$ at $z>2$). We have not replaced the photometric redshifts of those galaxies with spectroscopic redshifts. Given the very small percentage of $z>2$ sources with spectroscopic information, the impact of not introducing this change is negligible in all our analysis. We find a negligible bias ($\equiv \mathrm{median}(z_\mathrm{phot} - z_\mathrm{spec}$)) in our photometric redshifts. All galaxies lying outside the gray shaded area are considered outliers, which are defined as those sources for which $\sigma_i = |z_\mathrm{phot,i} - z_\mathrm{spec,i} | / (1 + z_\mathrm{spec,i}) > 0.15$. We find that only $\sim5 \%$ of sources are outliers according to this criterion and the remaining ones show a tight $z_\mathrm{phot}$-$z_\mathrm{spec}$ correlation, i.e. $\sigma_z = \mathrm{std}(\sigma_i) = 0.026$. The negligible bias, small $\sigma_z$ and small outlier fraction ($\eta$) show that our photometric redshifts are of excellent quality. These values are broadly consistent with those obtained by other authors in the literature, although in most cases the number of galaxies with spectroscopic redshifts utilized in the literature diagnostic is less than a half than the number used here (except in \citet{lai16}, where the total number of considered sources with spectroscopic redshifts is comparable to ours). 	 Considering only the sources with $z_{\rm spec}>2$, the bias and $\sigma_z$ are still small (-0.089 and 0.032, respectively), but the fraction of outliers rises to $\sim 16\%$.

\subsection{Stellar masses}
\label{sec:stmass}

\begin{figure}
\plotone{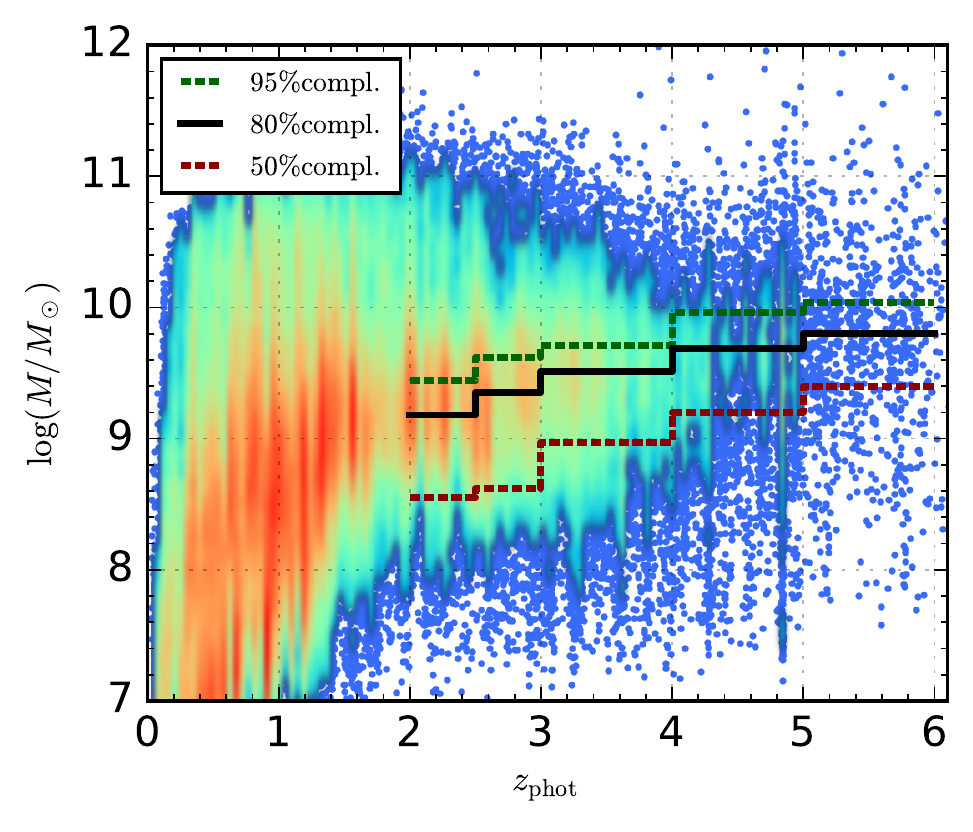}
\caption{Stellar masses vs. photometric redshifts for all SMUVS galaxies. The green dashed, solid black and red dashed lines show the 95\%, 80\% and 50\% completeness limits. For clarity, high (low) density regions in the plot have been coloured red (blue).}
\label{fig:stmvsz}
\end{figure}

Figure~\ref{fig:stmvsz} shows the best-fit stellar masses $M_*$  vs.  $z_{\rm phot}$ obtained from \textsc{LePhare}  for the SMUVS galaxies. This figure shows that our galaxy sample spans more than four decades in stellar mass, from $10^7$ through $>10^{11} \, \rm M_\odot$. To estimate the stellar-mass $50 \%$ completeness limits at different redshifts, we follow the method described in \citet{2014ApJ...783...85T}. First, we consider the IRAC limiting magnitude for which our sample can be considered 100\% complete, which is $m_\mathrm{[4.5],\, \rm lim} \leq$ 24.75 (see black dashed line in Fig. \ref{fig: completeness}), as discusssed in  Section~\ref{sec:data}.  Then we divided our sample in redshift bins of width $\Delta z=0.5$ and, in each of them, we work out the limiting stellar mass ($M_{*, \mathrm{lim}}$) at which 50\% of galaxies with $M_{*} >M_{*, \mathrm{lim}}$ have $m_{[4.5]}  <m_\mathrm{[4.5],\, \rm lim}$. Similarly, we also estimate the stellar-mass $80 \%$ and $95 \%$ completeness limits at different redshifts. We summarize our results in  Table~\ref{tab:masscompl}.

\begin{table}[h!]
\centering
\begin{tabular}{c|c|c|c}
        	& 50\% $M_*$ limit 	& 80\%  $M_*$ limit & 95\%  $M_*$ limit \\
Redshift	& $\log_{10} (M/M_\odot)$	& $\log_{10} (M/M_\odot)$  & $\log_{10} (M/M_\odot)$ \\ \hline \hline
2.0 - 2.5	& 8.55				& 9.18  &  9.44 \\
2.5 - 3.0	& 8.62				& 9.35  &  9.62 \\
3.0 - 4.0	& 8.97				& 9.51  &  9.71 \\
4.0 - 5.0	& 9.20				& 9.69  &  9.96 \\
5.0 - 6.0	& 9.40				& 9.80  &  10.04
\end{tabular}
\caption{Stellar-mass 50\%, 80\%  and 95\% completeness limits at different redshifts.}
\label{tab:masscompl}
\end{table}

\subsection{Selection of Dusty and Non-Dusty Galaxies}

\begin{figure}
\plotone{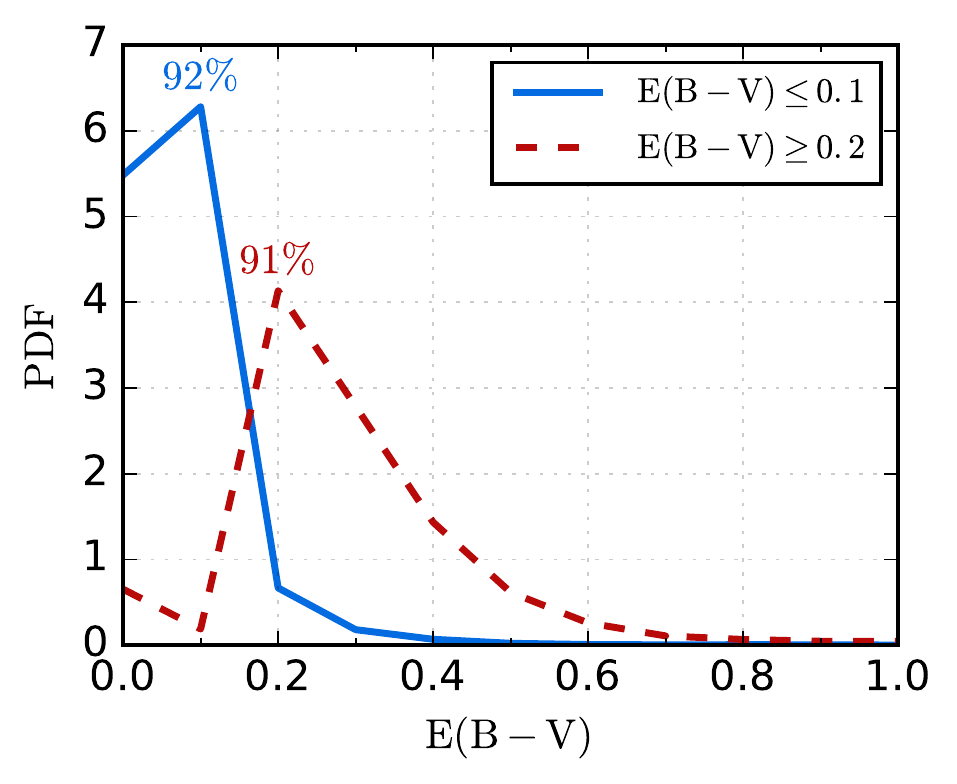}
\caption{Total probability density distribution versus colour excess $\rm E(B-V)$ for a representative sample of our dusty and non-dusty galaxies at $z=2-6$. These probability density distributions have been obtained by marginalizing over all other variables. The integrated PDF indicate that dusty (non-dusty) galaxies have an overall probability of 0.91 (0.92) of being classified in the correct extinction group.}
\label{fig:ebvpdf}
\end{figure}

The main goal of this paper is to analyse the stellar mass and redshift evolution of dusty and non-dusty galaxies at $z=2-6$.  Our classification is based on the colour excess $\rm E(B-V)$ that we obtain from the best-fit SED fitting: we divide our galaxy sample into two groups, one with $\rm E(B-V)\leq 0.1$  and another one with $\rm E(B-V)\geq 0.2$. These values correspond to $V$-band extinctions $A_V \lsim 0.4$~mag and $A_V \gsim 0.8$~mag, respectively, for a \citet{cal00} reddening law. We chose these colour excess values to divide the sample such that we have roughly similar numbers of galaxies in the two extinction groups. Across $z=2-6$ the overall median percentage of non-dusty galaxies varies between 40 and 70\%, depending on the redshift. Note that, according to our criterion,  non-dusty galaxies are indeed virtually dust-free, while the dusty group comprises galaxies with moderate to high dust extinctions.  This classification in dusty and non-dusty galaxies is slightly different to that considered by \citet{mar16}, who adopted an empirical division in the UVJ colour-colour diagram which approximately coincides with a dust extinction $A_V=1$. 

Our classification of dusty and non-dusty galaxies is robust against degeneracies in parameter space. Figure~\ref{fig:ebvpdf} shows the total probability density distribution versus colour excess $\rm E(B-V)$ for a representative sample of 800 dusty and non-dusty galaxies in our sample at $z=2-6$. These probability density distributions are obtained by marginalizing over all other variables. This figure shows that, even considering degeneracies in parameters space, the dusty and non-dusty galaxies have an overall probability $>0.9$ of being in their correct classification group.

\begin{figure*}
\centering
\includegraphics[width=\textwidth]{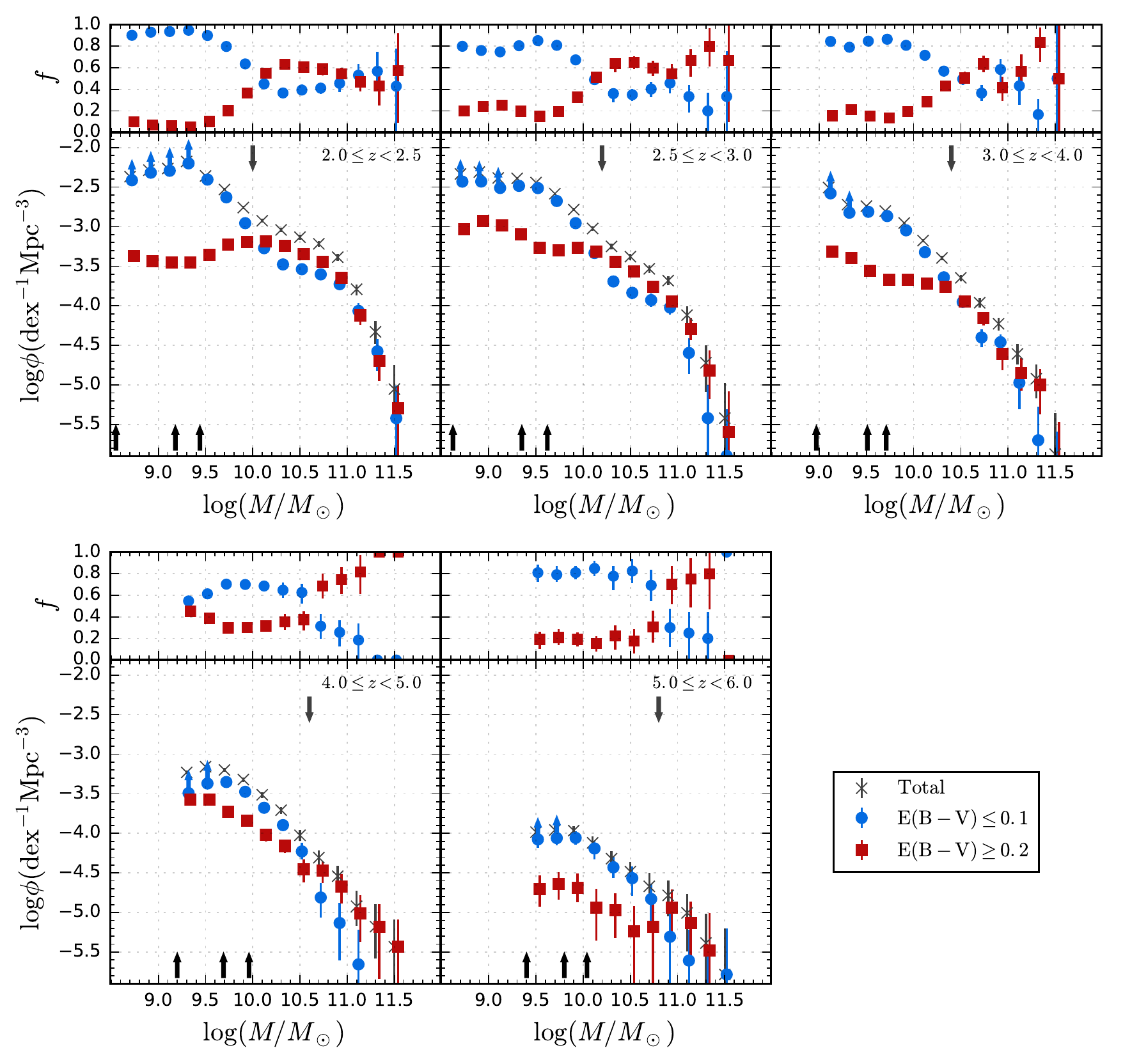}
\caption{The GSMF data points computed with $1/V_\mathrm{max}$ technique decomposed into two extinction bins. Error bars include Poisson shot noise, SED-fitting uncertainties, and cosmic variance. The total GSMF is shown with black crosses, the non-dusty and dusty GSMFs are shown with blue filled circles and red squares, respectively. The upper panels show the fraction $f$ of non-dusty and dusty galaxies with respect to the total GSMF. Black upwards pointing arrows indicate the 50\%, 80\% and 95\% stellar-mass completeness limits. Upwards pointing arrow on the data points indicate bins affected by sources fainter than IRAC=26. The downwards pointing arrow shows the limit at which we recover the high-mass end when considering IRAC $< 23.5$ sources. Errors on $f$ are determined by considering the variance of a binomial distribution (after incorporating non-Poissonian uncertainties).}
\label{fig:ext+frac}
\end{figure*}

\section{The GSMF of Dusty and Non-Dusty Galaxies at $z=2-6$} \label{sec:gsmf}

The GSMF at high $z$ has recently been studied in the COSMOS field \citep{ilb13,muz13,cap15,dav17}, and other fields \citep[e.g.,][]{cap11,san12,dun14,gra15,son16}. Here we focus on analysing the contributions of dusty and non-dusty galaxies to the GSMF at $z=2-6$. In the following we present our results for the two considered galaxy families. As a sanity check, we verify that our total GSMF is consistent with previous works at different redshifts: we show the results of this general comparison in Appendix~\ref{sec:gsmfcomparison}.

\subsection{Methodology}
\label{sec:methodology}
	
We compute the GSMF using the $1/V_\mathrm{max}$ technique \citep{sch68} at different redshifts. Although this technique involves binning the galaxy sample in stellar mass,  it has the advantage of being free of any parameter dependence or model assumptions. To compute $V_\mathrm{max}$, we need to calculate the maximum redshift $z_\mathrm{max}$ at which a source could have been observed given a limiting flux. This is accomplished by solving the following equation for $z_\mathrm{max}$:

\begin{equation}
	\frac{ D^2_\mathrm{lum}(z_\mathrm{max}) }{ 1+z_\mathrm{max} } = \frac{f_{\nu, \mathrm{obs} }}{ f_{\nu, \mathrm{lim} } } \frac{ D^2_\mathrm{lum}(z_\mathrm{obs}) }{ 1+z_\mathrm{obs} }
\end{equation}
where $D_\mathrm{lum}(z)$ is the luminosity distance at redshift $z$, and $f_{\nu, \mathrm{obs} }$, $f_{\nu, \mathrm{lim} }$ are the observed and limiting fluxes respectively. We choose a limiting flux corresponding to a magnitude [4.5]=26~mag  (or [3.6]=26~mag, in the case of non-detection at $4.5 \, \rm \mu m$). 
 %We computed the $V_\mathrm{max}$ correction for each galaxy considering the maximum volume at which the galaxy would have a magnitude [4.5]=26  (or [3.6]=26~mag, in the case of non-detection at $4.5 \, \rm \mu m$). 
For sources fainter than this limiting magnitude, we apply no $V_\mathrm{max}$ corrections. In addition to the $V_\mathrm{max}$ correction for each galaxy,  we apply an incompleteness correction factor ($100\%/x\%$) considering the [4.5] magnitude of each galaxy (or its [3.6] magnitude in the case of non detection at $4.5 \, \rm \mu m$) and the completeness levels ($x\%$) determined from the \cite{2012ApJ...752..113H} model (black dashed line) in Fig.~\ref{fig: completeness}.  
	
We identify three sources of uncertainties in the GSMF calculation, namely a Poisson error $\sigma_\mathrm{poi}$,  an error associated with the SED fitting $\sigma_\mathrm{mc}$, and cosmic variance $\sigma_\mathrm{cov}$. The first one is simply related to the statistics of our galaxy sample. We estimate $\sigma_\mathrm{poi}$ using the tabulated values provided in \citet{geh86}. The SED-fitting error is related to the uncertainties in the photometric redshifts and stellar-mass determinations.  To estimate $\sigma_\mathrm{mc}$,  we create 100 mock catalogs. These mock catalogs are obtained by randomizing the photometry of each galaxy (within the photometric uncertainties assuming a Gaussian distribution) and re-determining the masses and redshifts with \textsc{LePhare}. We then recompute the GSMF for each of the mock catalogs and $\sigma_\mathrm{mc}$ is the 16th and 84th percentiles of these mock GSMFs. %In these catalogs, the mock redshift of each galaxy has been derived randomly from the individual P(z) of each galaxy, as provided by \textsc{LePhare}. The corresponding mock stellar mass has been obtained randomly from a Gaussian distribution centred at the re-scaled original stellar mass (the re-scaling has been done taking into account the new, mock redshift). This Gaussian r.m.s. $\sigma_M$ has been obtained from the  $16^\mathrm{th}$ and $84^\mathrm{th}$ percentiles of the integrated PDF given by \textsc{LePhare}, at the best-fit $z_{\rm phot}$ value. 
Finally, to estimate the errors due to cosmic variance we followed the prescription of \citet{2011ApJ...731..113M}. We comment on the contribution of each source of uncertainty in Appendix \ref{sec: gsmf unc}.

Note that throughout this paper we only show the GSMF data points down to our estimated 50\% stellar-mass completeness limits (Table \ref{tab:masscompl}). And, as explained above, in the GSMF calculation we have considered all the SMUVS galaxies that result in stellar masses down to these limits (independently of their IRAC magnitudes). This allows us to show the widest possible dynamic range in stellar mass enabled by our data. We have performed a few sanity checks and confirmed the following: 1) if we only consider those galaxies with IRAC $< 23.5$, that is brighter than the 95\% completeness limit 
%within 95\% completeness, 
we recover the high-mass end of our GSMF (this is shown as a downward facing arrow in our GSMF plots; 2) if we exclude all sources with IRAC mag$>26$ (i.e. those sources below the 50\% completeness limit of the IRAC catalogue), we basically obtain the same GSMF as that shown here. Only in the lowest stellar-mass bins do we observe some marginal difference, which is irrelevant, as no conclusion in this paper depends on them. The GSMF data points which include sources with mag$>26$ are indicated as lower limits in our GSMF plots.

\begin{figure}
\plotone{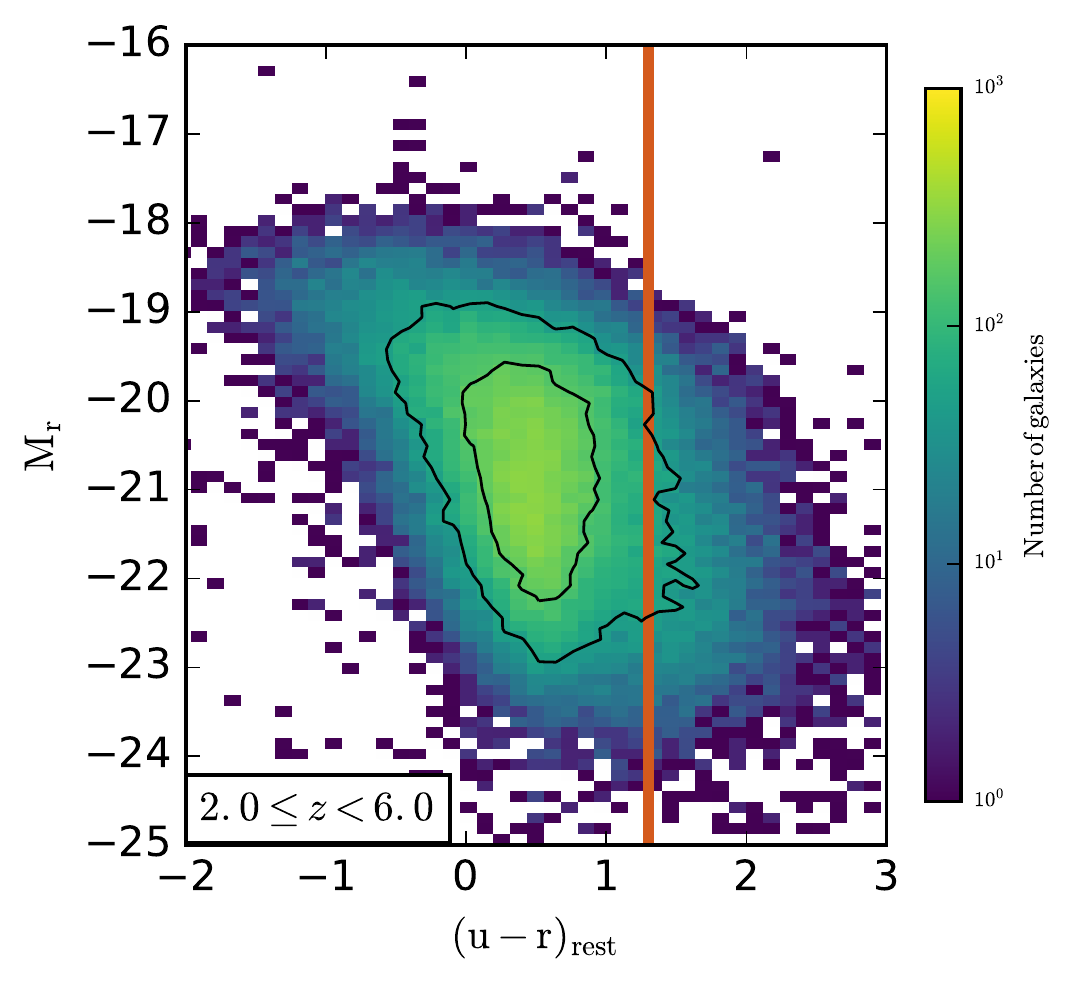}
\caption{Rest-frame colour magnitude diagram of our sources at $2.0 < z < 6.0$. The contours contain the 50 and 80\% of our galaxies. The vertical red line corresponds to our colour cut \citep{bal04}. The rest-frame colours are derived using the filters closest to rest-frame $u$ and $r$.}
\label{fig:colormag}
\end{figure}

\begin{figure*}
\centering
\includegraphics[width=\textwidth]{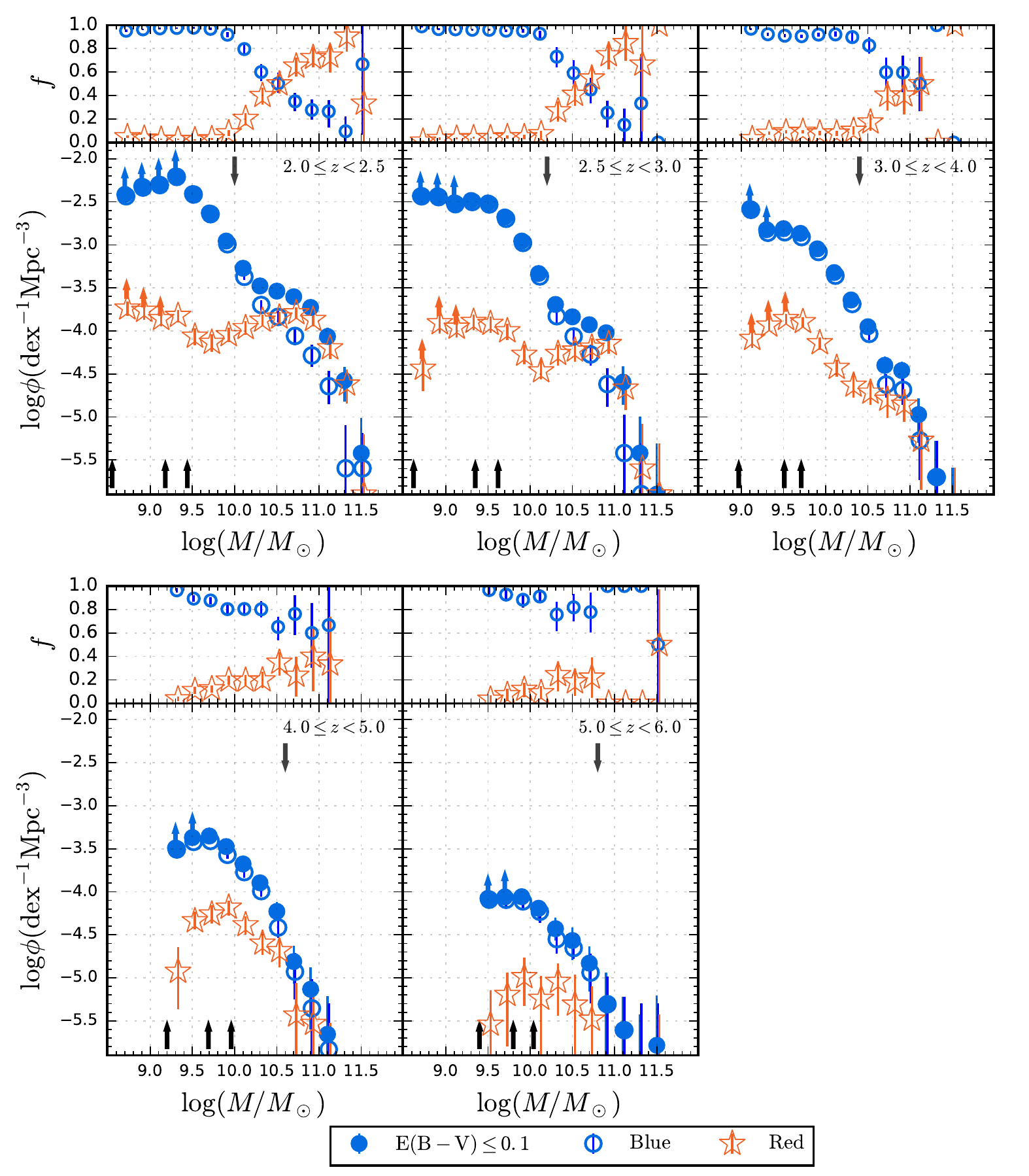}
\caption{Decomposition of the non-dusty GSMF for blue and red galaxies. We show the total non-dusty GSMF (blue filled circles) for reference, as well as the GSMF of the blue and red populations (blue open circles and red stars respectively). The upper panels show the fraction of blue and red galaxies in the non-dusty sample. Black upwards pointing arrows indicate the 50\%, 80\% and 95\% stellar-mass completeness limits. Upwards pointing arrow on the data points indicate bins affected by sources fainter than IRAC=26. The downwards pointing arrow shows the limit at which we recover the high-mass end when considering IRAC $< 23.5$ sources.}
\label{fig:dfcolors}
\end{figure*}

\subsection{General Results}
\label{sec:gsmfresults}

Figure~\ref{fig:ext+frac} shows our GSMF computed with the $1/V_\mathrm{max}$ method and corrected for completeness for the dusty and non-dusty galaxies separately, at different redshifts from $z=2$ to $z=6$. On top of each panel, we show  the fraction $f$ of the two different populations as a function of stellar mass. We calculate the uncertainties in $f$ considering a binomial distribution. The 50\%, 80\% and 95\% stellar-mass completeness limits at different redshifts are indicated with black upwards pointing arrows in the GSMF panels. All GSMF values for dusty, non-dusty, non-dusty blue and non-dusty red galaxies are tabulated in Appendix \ref{sec: gsmf tables}.

From Fig.~\ref{fig:ext+frac} we can see that, at $z=2.0-2.5$, dusty and non-dusty galaxies contribute similarly to the overall population of galaxies with stellar masses $\gsim 10^{10.1} \, \rm M_\odot$. At lower stellar masses, instead, the GSMF is clearly dominated by the non-dusty galaxies.

At $z>2.5$ dusty galaxies start to dominate the GSMF high-mass end, or become comparable in number density to non-dusty galaxies, making for 60\% to 80\% of all massive galaxies at these redshifts. The dusty galaxy dominance becomes most evident at $4<z<5$.  The stellar mass below which non-dusty galaxies overtake the dominance evolves with redshift: it is  $\approx 10^{10.5} \, \rm M_\odot$ at $z=3.0-4.0$ (compared to $\approx 10^{10.1} \, \rm M_\odot$ at $z=2.5-3.0$).

In addition, our results indicate that the period elapsed at $z=4-5$ was of major importance for dust extinction in galaxy evolution.  Dusty galaxies more clearly dominate the GSMF high-mass end at $M_* \gsim 10^{10.6} \, \rm M_\odot$, and their fraction increases steadily with stellar mass, reaching $>80\%$ at $M_* >10^{11} \, \rm M_\odot$. Below $M_* \sim 10^{10.6} \, \rm M_\odot$, non-dusty galaxies are more numerous than dusty ones, following the same trend observed at lower redshifts. However, at $z=4-5$ the percentage of dusty sources among intermediate-mass galaxies is higher than at any later time, i.e.,  30-40\% of all intermediate and low-mass galaxies down to $M \approx 10^{9} \, \rm M_\odot$. This indicates that significant dust extinction was not only important among massive galaxies, but also among many lower mass galaxies at these redshifts.

In the total GSMF at $z=2.0-2.5$  (Fig.~\ref{fig:ext+frac} and Appendix~\ref{sec:gsmfcomparison}),  there is a flat regime at intermediate stellar masses. This feature has previously been identified in the literature at lower redshifts \citep[e.g.,][]{poz10,bie12}, implying that the GSMF is best fit by a double \citet{sch76} function rather than a single one. Interestingly, this kind of double functional form is more clearly seen for the non-dusty galaxies alone, with our results indicating that this is a feature present since at least $z=3$.

\begin{figure*}
\centering
\includegraphics[width=\textwidth]{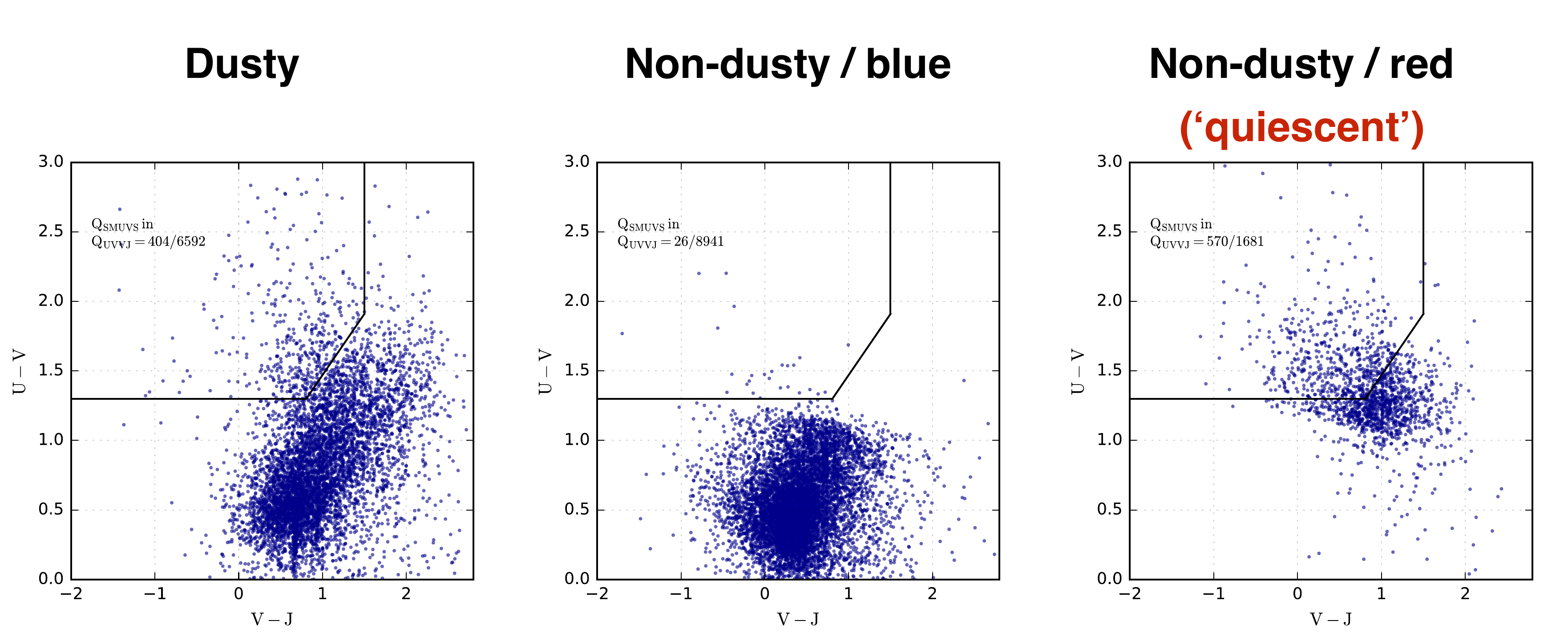}
\caption{Location of our three classified galaxy populations (dusty; non-dusty/blue; and non-dusty/red) in the rest $UVJ$ colour-colour diagram, showing the wedge utilized by other authors to segregate `quiescent galaxies'. For clarity, we only show our galaxies with $\log_{10} M_*/M_\odot \geq 9.7$, which are those that are the main focus of discussion in this paper. These plots show that only about a third of our classified quiescent galaxies lie within the $UVJ$ quiescent wedge. We recognise many more quiescent galaxies that are close to the wedge, but lie outside of it (right panel). At the same time, we find that the quiescent colour wedge has significant contamination from dusty sources (left panel). Our non-dusty/blue sources virtually all lie outside of the wedge, as expected (middle panel).}
\label{fig:uvj}
\end{figure*}

\subsection{Bisection of the non-dusty galaxy population}

Non-dusty objects can be very diverse in nature: some are unobscured star-forming galaxies, while others have no dust because they have virtually ceased their star formation, i.e., they are old and passive (or almost passive) galaxies. These two groups can broadely be divided using rest-frame optical colours, as it is commonly done in the literature (although in general these colour criteria are applied without separating dusty and non-dusty galaxies in the first place).

Fig.~\ref{fig:colormag} shows the rest-frame  $M_r$ absolute magnitude vs. $u-r$ colour diagram  for our non-dusty galaxies at $z=2.0-6.0$.	 We compute the absolute magnitudes by applying a $k$-correction to the filters closest to the rest-frame $u$ and $r$ bands. This method is preferred to the one where the absolute magnitudes are computed directly from SED templates (see Section 4.3 in \citet{dav17} and references therein for a detailed discussion). We then classify our non-dusty galaxies  according to their colours:  $u-r < 1.3$ and $u-r \geq 1.3$,  as the blue and red populations, respectively, following \citet{bal04}.

We find that only $\sim 6\%$ of our non-dusty galaxies are quiescent, i.e., have a red $u-r$ colour. As these galaxies are non-dusty, their red optical colours can only be explained by the presence of a prominent $\rm 4000 \, \rm \AA$ break, i.e. they are old galaxies, dominated by stars with ages $\gsim 1 \, \rm Gyr$.  Perhaps not surprisingly, this minor fraction of red, non-dusty sources are very biased in stellar mass: they are mostly massive galaxies, as can be seen in Fig.~\ref{fig:dfcolors}. We will discuss massive galaxies further in Section~\ref{sec:massivegal}.

To check for possible contaminants among our classified quiescent galaxies, we have cross-correlated this population with the {\em Spitzer} COSMOS $24 \, \rm \mu m$ catalogue \citep{san07} and the C-COSMOS X-ray catalogue \citep{2016ApJ...819...62C}. We found that only 1.5\% of our classified passive sources are X-ray AGN, and only 3\% are $24 \, \rm \mu m$ detected, indicating that the fraction of contaminants within our sample is very small.

Almost all our dusty and non-dusty red galaxies $(97-99\%)$ are brighter than IRAC=26. On the other hand, non-dusty blue sources constitute $\sim96\%$ of the IRAC-faint population (IRAC$>26$).

The bisection of the non-dusty GSMF into red and blue sources clearly shows the origin of the double-Schechter behaviour up to $z=3$: while the low-mass regime is dominated only by blue (very likely star-forming) galaxies, the high-mass regime is made of both blue star-forming galaxies and red old galaxies. This is consistent with what has been found in the literature at lower redshifts \citep[e.g.,][]{poz10,ilb13}. More recently, \citet{2014ApJ...783...85T} have found a similar result up to $z=2$, but did not find an `upturn' in the GSMF at higher redshifts, in spite of analysing sufficiently deep data to investigate the relevant stellar-mass regime. Here, instead, we clearly see this upturn up to $z=3$ and confirm that this feature is present independently of any $V_\mathrm{max}$ and incompleteness corrections in our GSMF.

To facilitate the comparison with the galaxy population classification based solely on the rest $UVJ$ colour-colour diagram adopted by other authors \citep[e.g.,][]{muz13,2014ApJ...783...85T}, we show the locus that each of our classified galaxy populations (dusty; non-dusty/blue; and non-dusty/red) occupy on that plane (Fig. \ref{fig:uvj}). For clarity, we only show our galaxies with $\log_{10} M_*/M_\odot \geq 9.7$, which are those that are our main focus of discussion hereafter.

We find that only about a third of our classified quiescent galaxies lie within the quiescent wedge defined in the literature on the $UVJ$ plane. Within our own classification, we recognise many more quiescent galaxies that are close to the wedge, but lie outside of it. At the same time, we find that the quiescent colour wedge has significant contamination from dusty sources. Even if these sources can be relatively old (as much as the age of the Universe at each given redshift), the presence of significant dust extinction at high $z$ excludes the possibility that these galaxies can be strictly passive: a significant amount of dust extinction in the SED fitting implies that there is a significant amount of intrinsic UV photons, which can only be explained through significant star formation or nuclear activity. In summary,  if we applied a simple $UVJ$ colour-colour selection to classify quiescent galaxies, we would select a population which, according to our SED fitting results based on 28-band information, has $\sim 40\%$ of non-quiescent galaxies, and at the same time misses two-thirds of them (according to our quiescent definition based on SED fitting adopted here). So we can conclude that the two methodologies would select two different sub-samples of `quiescent galaxies'. Through the UVJ diagram one can mainly select galaxies with low specific star formation rates (e.g., \cite{2014ApJ...796...35F}), while our own methodology selects a population of galaxies for which the red rest-frame ($u-r$) colours are only explained by age rather than dust, i.e. galaxies that are more settled into a `red and dead' phase. 

Of course, one could argue that for many sources lying around the borders of the $UVJ$ passive wedge, the typical errors in the photometry (which are $\gsim 0.15$ for all ground-based colours) could explain the different classification. In fact, this the main reason why one should consider the alternative approach proposed here to select quiescent galaxies: by taking into account the entire photometric information, %In fact, this is another important reason to prefer our own approach rather than a simple colour-colour segregation: by taking into account the entire photometric information, 
the full SED fitting classification can compensate for the errors that may affect a particular colour.

\section{The Overall Fractions of Dusty/Non-Dusty Galaxies and the Evolution to Quiescence}
\label{sec:evolquies}

\subsection{Massive galaxies}
\label{sec:massivegal}

\begin{figure*}
\plotone{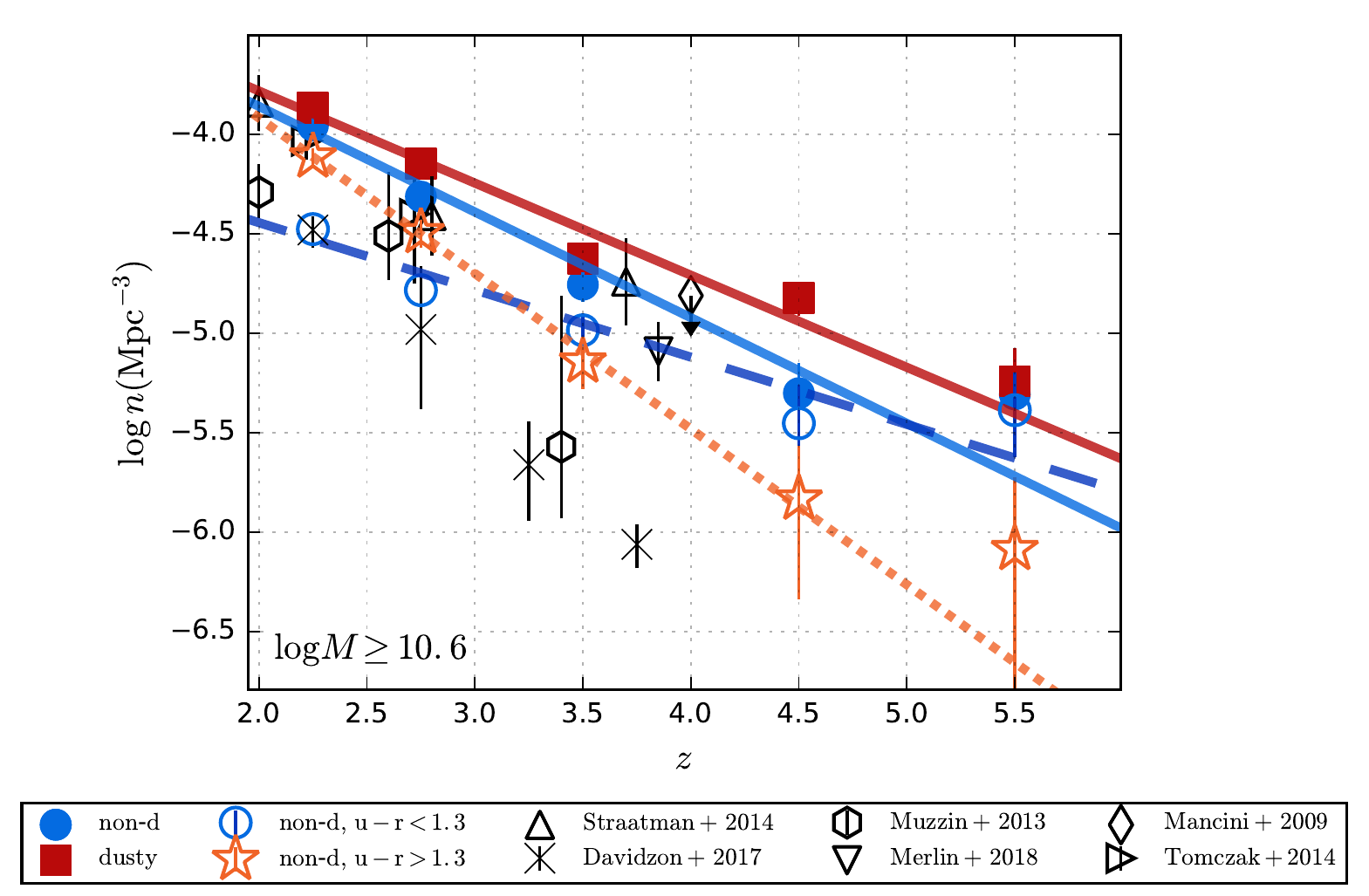}
\caption{Number densities $n$ of massive galaxies (i.e.,  those with $\log_{10} M_{\rm st}> 10.6 \, \rm M_\odot$) versus redshift. We show $n$ separately for different galaxy populations: dusty (squares), all non-dusty (filled circles), red non-dusty (stars) and blue non-dusty (open circles).  We show linear fits for each population. For a comparison, we plot the number densities of quiescent galaxies derived by \cite{man09}, \cite{muz13}, \cite{str14}, \citet{2014ApJ...783...85T}, \citet{dav17}, and \cite{mer18}.}
\label{fig:massiveevol}
\end{figure*}

In this Section we focus on the analysis of the population of massive galaxies with $\log_{10} M_*/M_\odot \geq 10.6$. Fig.~\ref{fig:massiveevol} shows the cumulative number density of galaxies with stellar masses above this threshold versus redshift. As before, we separate our sample in dusty and non-dusty galaxies, and the latter group was further divided according to their optical colours. Our derived number densities are also listed in Table~\ref{tab:massiveevol}.

We see that, at $z=4-5$, the population of massive galaxies is dominated by galaxies with high dust extinction. At $z\sim 2$
the total number density of massive galaxies is more than a factor of ten larger than at $z=4-5$, and the dusty and non-dusty populations become similar in number density.  By analysing the number density evolution, we infer that a possible explanation for this behaviour is that the dusty massive galaxies at $z\sim 4-5$ evolve into non-dusty sources by $z\sim3.5$ (both number densities are very similar), while at the same time new dusty ones are being created. Similarly, the dusty galaxies at $z\sim3.5$ could become non-dusty by $z\sim 3$, while further dusty ones are created,  and this process continues at least down to $z=2$.

Among the non-dusty galaxies, the blue and red populations evolve at different rates with cosmic time. While the number density of blue non-dusty galaxies grows by one dex between $z\sim5.5$ and $z\sim2$, the number density of red non-dusty galaxies increases by two dex in this same period. Our red non-dusty galaxies are, by definition, a very good proxy for quiescent galaxies\footnote{strictly, determining the passive nature requires spectroscopy to confirm the absence of significant emission lines. However, our selection criterion of being red and non-dusty naturally selects galaxies dominated by old stellar populations. In this sense, our classification is better than selections based only on colour cuts to identify quiescent galaxies, which as we showed have significant contamination from dusty sources.}. Therefore, this means that the population of massive quiescent galaxies has grown by a factor of $\sim 100$ in the $\sim$2~Gyr elapsed between $z=5.5$ and $z=2$.

The number density evolution of blue and red non-dusty galaxies shown in Fig.~\ref{fig:massiveevol} suggests that it is unlikely that dusty galaxies could have evolved into red non-dusty (quiescent) galaxies directly. If there is an evolutionary link between these massive galaxy populations, then the most likely sequence is: dusty (star-forming) $\rightarrow$ blue non-dusty (star-forming) $\rightarrow$ red non-dusty (quiescent).  This would be consistent with a scenario in which massive galaxies passed by a non-dusty star-forming phase before becoming quiescent objects. A plausible physical mechanism for stripping massive galaxies of their dust and subsequently quenching star formation are high-velocity outflows driven by black hole accretion \citep[e.g.][]{hop16}. Our galaxy number densities suggests that this transition from dusty into blue non-dusty takes about 0.5-1.0~Gyr. In any case, although this proposed evolutionary scenario is compatible with our derived galaxy number densities, our results can certainly not exclude other possible paths for massive galaxy evolution.

In Fig.~\ref{fig:massiveevol} we also compare the number density of our red non-dusty massive galaxies with the number density of passive galaxies selected in the literature for the same (or a very similar) stellar mass cut \citep{man09,muz13,str14,2014ApJ...783...85T,dav17,mer18}. These passive galaxy selections are based on rest-frame colours and their number densities are directly quoted by the authors, or we have obtained them by integrating their corresponding GSMF\footnote{In the latter case we integrate the Schechter fits to the GSMF. The uncertainties on the number densities reflect the uncertainties on the Schechter parameters. This method was implemented for \cite{muz13,2014ApJ...783...85T} and \cite{dav17}. }. Our comparison shows that the number density of our red non-dusty galaxies are significantly higher than the number density of passive galaxies derived from \citet{muz13} and \citet{dav17} GSMFs. These differences are mainly produced by the shallower depths of the datasets used by these authors with respect to our own. In the case of the comparison with \citet{dav17}, there is the effect of (a) their choice of dust extinction law and (b) their specific criterion to select passive galaxies, which is colour based, but different to that applied by other authors. \cite{dav17} also correct for Eddington bias \citep{1913MNRAS..73..359E} which causes a drop in number densities at the massive end of the GSMF.

Our red non-dusty galaxy number densities are broadly consistent with the value derived by \citet{mer18} and upper limit from \cite{man09}.  They are also in excellent agreement with the number densities derived by \citet{2015ApJ...808L..29S} at $z\sim 2-3$, although this might be somewhat fortuitous, given their very different methodology to select quiescent galaxies.  At higher redshifts, instead,  our results indicate a much faster decline in the number density of passive galaxies than that obtained by these authors. This difference does not seem to be the product of the different methodologies (note that we would select {\em less} rather than more passive galaxies using the quiescent wedge of the $UVJ$ diagram, according to Fig.~\ref{fig:uvj}). The observed differences could in part be the result of cosmic variance, as \citet{str14} analyzed images over an area $\sim 6.5$ times smaller than that considered here.

\begin{table*}
\centering
\begin{tabular}{c|l|l|l|l}
Redshift	& $n_{\rm dusty}$ (all) & $n_{\rm non-dusty}$ (all) &  $n_{\rm non-dusty}$ (blue) & $n_{\rm non-dusty}$ (red) \\
& $10^{-4}\times \mathrm{Mpc^{-3}}$  & $10^{-4}\times \mathrm{Mpc^{-3}}$	& $10^{-4}\times \mathrm{Mpc^{-3}}$ &  $10^{-4}\times \mathrm{Mpc^{-3}}$ \\
\hline 
\hline
2.25  & $1.366^{+0.107}_{-0.107}$ & $1.107^{+0.101}_{-0.082}$  &  $0.335^{+0.053}_{-0.050}$ & $0.773^{+0.080}_{-0.068}$ \\ 
2.75 &  $0.713^{+0.077}_{-0.078}$  & $0.487^{+0.058}_{-0.071}$  & $0.165^{+0.041}_{-0.032}$ & $0.322^{+0.051}_{-0.054}$ \\
3.50  & $0.238^{+0.033}_{-0.033}$  & $0.176^{+0.029}_{-0.027}$  & $0.104^{+0.023}_{-0.021}$ & $0.072^{+0.020}_{-0.019}$ \\ 
4.50  & $0.151^{+0.031}_{-0.029}$ & $0.050^{+0.022}_{-0.017}$ & $0.035^{+0.020}_{-0.015}$ & $0.015^{+0.012}_{-0.010}$ \\
5.50  & $0.058^{+0.027}_{-0.028}$ & $0.049^{+0.025}_{-0.018}$ & $0.041^{+0.022}_{-0.017}$ & $0.008^{+0.012}_{-0.008}$
\end{tabular}
\caption{Number densities of massive ($\log_{10} M_*/M_\odot > 10.6$) galaxies.}
\label{tab:massiveevol}
\end{table*}

\begin{figure*}
\centering
\includegraphics[width=\textwidth]{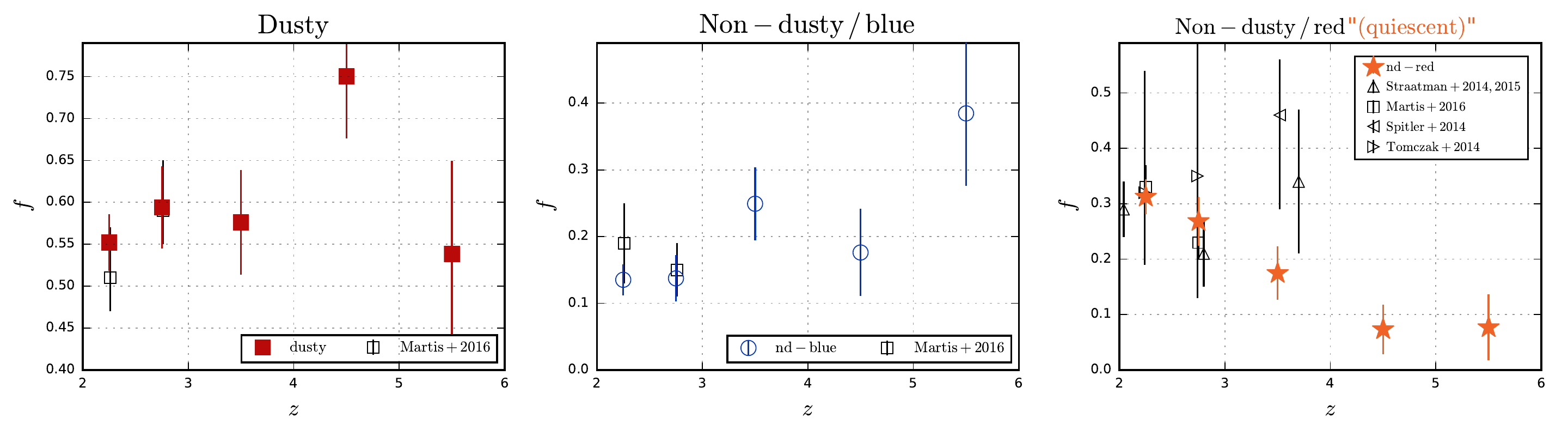}
\caption{Comparison of our fractions of massive ($\log_{10} M_*/M_\odot > 10.6$) galaxies with different classifications, with the fractions of similar populations from the literature. In contrast with our methodology, all literature works cited here have selected quiescent galaxies using the wedge in the $UVJ$ colour-colour diagram. The \cite{mar16} data points for dusty and non-dusty star-forming galaxies  are also based on regions empirically defined on this colour-colour plot, calibrated using the galaxy SED dust extinctions.}
\label{fig:fractions}
\end{figure*}

Fig.~\ref{fig:fractions} shows the fractions of our different populations (dusty, non-dusty/blue and non-dusty/red, i.e. quiescent) among all massive ($\log_{10} M_*/M_\odot > 10.6$) galaxies, versus redshift. We compare these fractions to those obtained in the literature. In these other works,  quiescent galaxies have been selected using the $UVJ$ colour-colour diagram wedge. In addition, \citet{mar16} have also determined regions to segregate dusty and non-dusty star-forming galaxies on this colour-colour plane.

We find that our fractions of massive galaxies classified in the different groups at $2<z<3$ are in good agreement, within the error bars,  with the fractions reported by \citet{mar16}. Our fraction of quiescent galaxies at these redshifts also broadely agrees with the fraction obtained from  \citet{2014ApJ...783...85T} and \citet{2015ApJ...808L..29S}, although the error bars in the fractions derived from these other works are very large.

%At $z>3$, our fraction of quiescent galaxies among massive galaxies is sginificantly lower than the values reported by \citet{str14, 2015ApJ...808L..29S} and \citet{2014ApJ...787L..36S} (only barely consistent with the very large error bars in \citet{str14}). Although these differences could be ascribed to the other authors using a different methodology to select quiescent galaxies, this is unlikely to be the main reason for this discrepancy. Note that despite \citet{str14} and \citet{2015ApJ...808L..29S} use the same technique to select quiescent galaxies, their results do not appear to be self-consistent, as the fraction of passive galaxies is higher at $z>3$ than at $z<3$. This would imply either that galaxies that are passive become non-passive at later cosmic times, or that the new massive galaxies present at $z<3$ cannot become quiescent, diluting the fraction of passive galaxies. Any of these scenarios seems implausible. In fact \citet{2015ApJ...808L..29S} acknowledge that there is a mismatch in their values. They propose that this could be due to dust-obscured star-forming galaxies contaminating their quiescent sample. 

\subsection{Intermediate-Mass Galaxies}
\label{sec:intermgal}

\begin{figure*}
\plotone{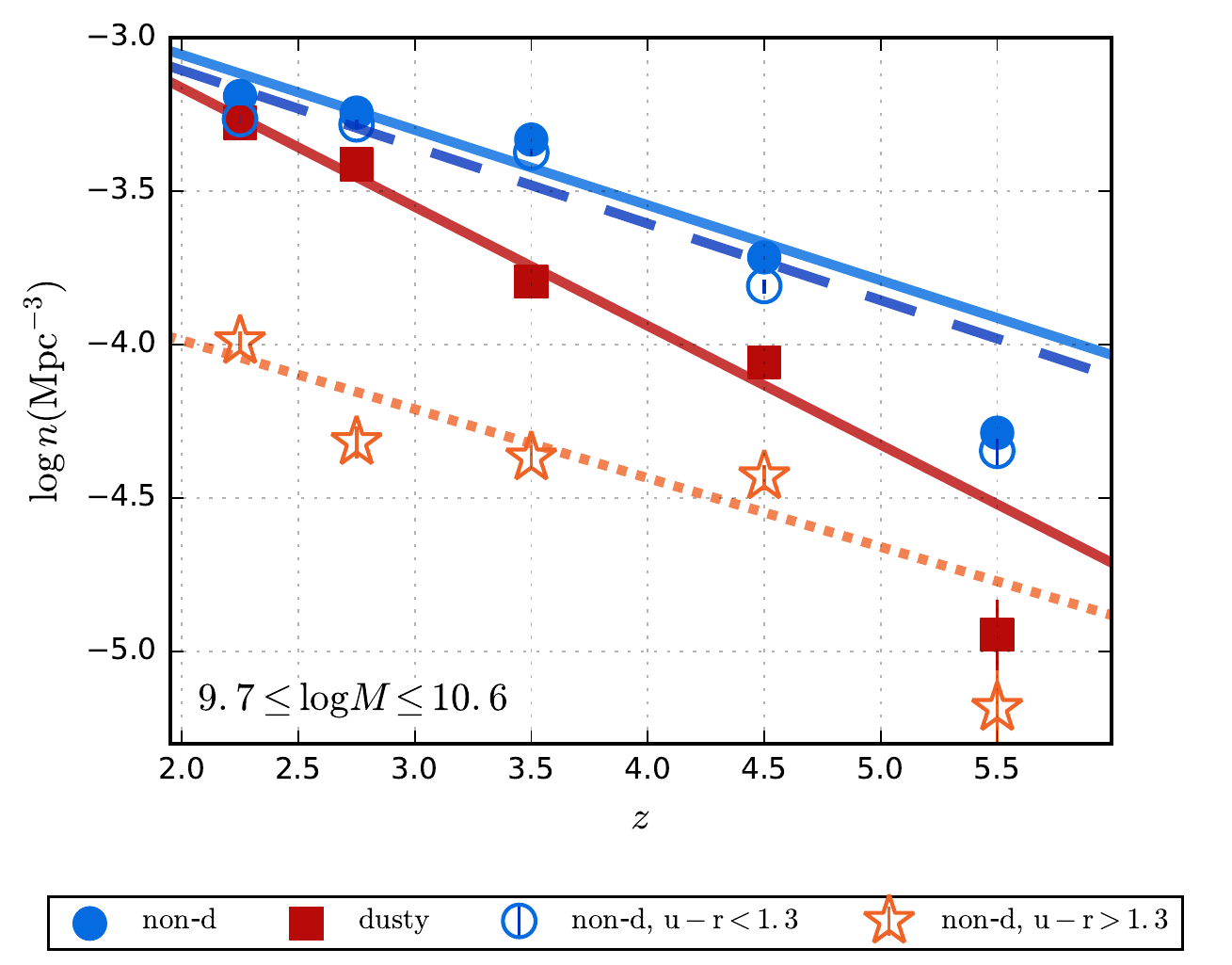}
\caption{Number densities $n$ of intermediate-mass galaxies (i.e.,  those with $9.7 < \log_{10} M_{\rm st} < 10.6 \, \rm M_\odot$) versus redshift. Symbols are the same as in Fig.~\ref{fig:massiveevol}.}
\label{fig:intermevol}
\end{figure*}

\begin{table*}
\centering
\begin{tabular}{c|l|l|l|l}
Redshift	& $n_{\rm dusty}$ (all) & $n_{\rm non-dusty}$ (all) &  $n_{\rm non-dusty}$ (blue) & $n_{\rm non-dusty}$ (red) \\
& $10^{-4}\times \mathrm{Mpc^{-3}}$  & $10^{-4}\times \mathrm{Mpc^{-3}}$	& $10^{-4}\times \mathrm{Mpc^{-3}}$ &  $10^{-4}\times \mathrm{Mpc^{-3}}$ \\
\hline 
\hline
2.25  & $5.287^{+0.198}_{-0.186}$ & $6.459^{+0.234}_{-0.183}$  & $5.433^{+0.208}_{-0.172}$ & $1.026^{+0.084}_{-0.079}$ \\ 
2.75 &  $3.865^{+0.169}_{-0.159}$  & $5.704^{+0.205}_{-0.197}$  & $5.223^{+0.190}_{-0.187}$ & $0.481^{+0.061}_{-0.068}$ \\
3.50  & $1.605^{+0.083}_{-0.077}$  & $4.659^{+0.120}_{-0.124}$  & $4.230^{+0.119}_{-0.123}$ & $0.428^{+0.043}_{-0.037}$ \\ 
4.50  & $0.875^{+0.059}_{-0.072}$  & $1.925^{+0.087}_{-0.096}$  & $1.553^{+0.078}_{-0.087}$ & $0.372^{+0.037}_{-0.038}$ \\
5.50  & $0.114^{+0.034}_{-0.029}$ &  $0.516^{+0.047}_{-0.057}$ & $0.451^{+0.044}_{-0.051}$ & $0.065^{+0.021}_{-0.017}$
\end{tabular}
\caption{Number densities of intermediate stellar-mass ($9.7 \leq \log_{10} M_*/M_\odot \leq 10.6$) galaxies.}
\label{tab:intermndens}
\end{table*}

Fig.~\ref{fig:intermevol} is analogous to Fig.~\ref{fig:massiveevol}, but for intermediate-mass galaxies. Here, we only analyse the stellar mass range $9.7 \leq \log_{10} M_*/M_\odot \leq 10.6$ to ensure high stellar mass completeness across all the analysed redshifts ($\gsim 80\%$ up to $z=5$, and $\sim 70\%$ completeness at $z=5-6$). In any case, all our quoted number densities carry completeness corrections, even if they are very small.  These derived number densities are listed in Table~\ref{tab:intermndens}.

We clearly see that the number density evolution of dusty and non-dusty intermediate-mass galaxies is very different to that of massive galaxies. At all redshifts $z>2.5$, blue non-dusty galaxies dominate the population with intermediate stellar masses. Only at $z=2$ does the population of dusty sources become equally important (in number) to the blue non-dusty galaxies. This is in contrast to the fractions observed among massive galaxies, in which the balance between dusty/non-dusty sources is fairer at $z=2-4$, and dominated by dusty galaxies at $z>4$. These results are consistent with the conclusions of \citet{mar16}, who found that intermediate stellar-mass galaxies are predominantly unobscured star-forming objects at $z\sim3$, while the high-mass galaxy population is dominated by dusty star-forming sources.

Another striking difference for intermediate-mass galaxies is that the percentage of red non-dusty (i.e., quiescent) sources is low and almost constant at all redshifts. Indeed, we obtain that quiescent sources constitute only $\sim 10\%$ of these galaxies at $z=2-6$, which suggests that star-formation quenching and evolution into quiescence is a much slower process among intermediate-mass galaxies than among massive ones. Our findings are consistent with the results of \citet{som14}, who reported a decline in the fraction of quiescent galaxies at faint near-IR magnitudes, corresponding to stellar masses $\lsim 10^{10.8} \, \rm M_\odot$.

\section{Summary and Conclusions}
\label{sec:conclus}
	
We have studied the evolution of dusty and non-dusty galaxies with stellar mass at $z=2-6$, considering the $\sim$66,000 SMUVS sources present in this redshift range. We classified our galaxies into dusty/non-dusty according to their colour excess E(B-V), as obtained with the best SED fitting solution. Furthermore, we divided the non-dusty sample using rest-frame optical colours to isolate the sample of quiescent galaxies (here defined as those galaxies whose red colours can only be explained by the dominance of old stellar populations).

For an overall statistical analysis, we computed the GSMF of our galaxy samples in different redshift bins between $z=2$ and $z=6$. We found that, at $z=2.0-2.5$, dusty and non-dusty galaxies contribute similarly to the overall population of galaxies with stellar masses $\gsim 10^{10.1} \, \rm M_\odot$. At $z>2.5$, instead, dusty galaxies dominate the GSMF high-mass end, making for 60\% to 80\% of all massive galaxies. The stellar mass below which non-dusty galaxies dominate evolves with redshift: it is  $\approx 10^{10.5}$ ($10^{10.1}$)~$\rm M_\odot$ at $z=3.0-4.0$ ($z=2.5-3.0$). The increasing importance of dust extinction with stellar mass is in agreement with the results of previous studies \citep{2012ApJ...754...25R, 2013MNRAS.429.1113H}.

At lower stellar masses the GSMF is clearly dominated by the non-dusty galaxies. At all the analysed redshifts, except at $z=4-5$, non-dusty galaxies make for $\sim 80\%$ of intermediate-mass galaxies. At $z=4-5$, instead, this percentage is somewhat lower, i.e. $\sim 60-70\%$. At this cosmic epoch, dusty galaxies appear to be at the maximum of their importance: they constitute 30-40\% of
the galaxies with  $M_*=10^9 - 10^{10.5} \, \rm M_\odot$ and  $>80\%$ of those with  $M_*>10^{11} \, \rm M_\odot$.

We also analyzed the evolution of quiescent galaxies among massive and intermediate-mass galaxies  (with $\log_{10} (M_*/M_\odot)\geq10.6$ and $\log_{10} (M_*/M_\odot)=9.7-10.6$, respectively) versus cosmic time. We found that the fraction of passive galaxies had a fast increase between $z\sim 6$ and $z\sim2$,  rising from $<10\%$ to $\sim 30\%$, which indicates that the mechanisms that quenched the star-formation activity among massive galaxies were very effective in the first few billion years of cosmic time. In remarkable contrast, the quiescent galaxy percentage among intermediate-mass galaxies stays rather constant at a $\sim 10\%$ level in this redshift range. These results are in line with the idea of {\em galaxy downsizing} \citep[e.g.,][]{be00,kod04,jun05,cat08}, and show that massive and intermediate-mass galaxies clearly had different evolutionary paths over the first few billion years of cosmic time.

\acknowledgments

Based in part on observations carried out with the {\em Spitzer Space Telescope}, which is operated by the Jet Propulsion Laboratory, California Institute of Technology under a contract with NASA. Also based on data products from observations conducted with ESO Telescopes at the Paranal Observatory under ESO program ID 179.A-2005 and on data products produced by TERAPIX and the Cambridge Astronomy Survey Unit on behalf of the UltraVISTA consortium. Also based on observations carried out by NASA/ESA {\em Hubble Space Telescope}, obtained and archived at the Space Telescope Science Institute; and the Subaru Telescope, which is operated by the National Astronomical Observatory of Japan.  This research has made use of the NASA/IPAC Infrared Science Archive, which is operated by the Jet Propulsion Laboratory, California Institute of Technology, under contract with NASA.

KIC, SD and WIC  acknowledge funding from the European Research Council through the award of the Consolidator Grant ID 681627-BUILDUP.

The Cosmic Dawn center is funded by the DNRF.


\vspace{5mm}
\facilities{Spitzer, VISTA, Subaru}


\software{SExtractor, IRAF, LePhare}





\appendix

\section{Photometry comparison with public COSMOS catalogues}	
\label{sec:photcomparison}

\begin{figure*}
\centering
\includegraphics[width=0.95\textwidth]{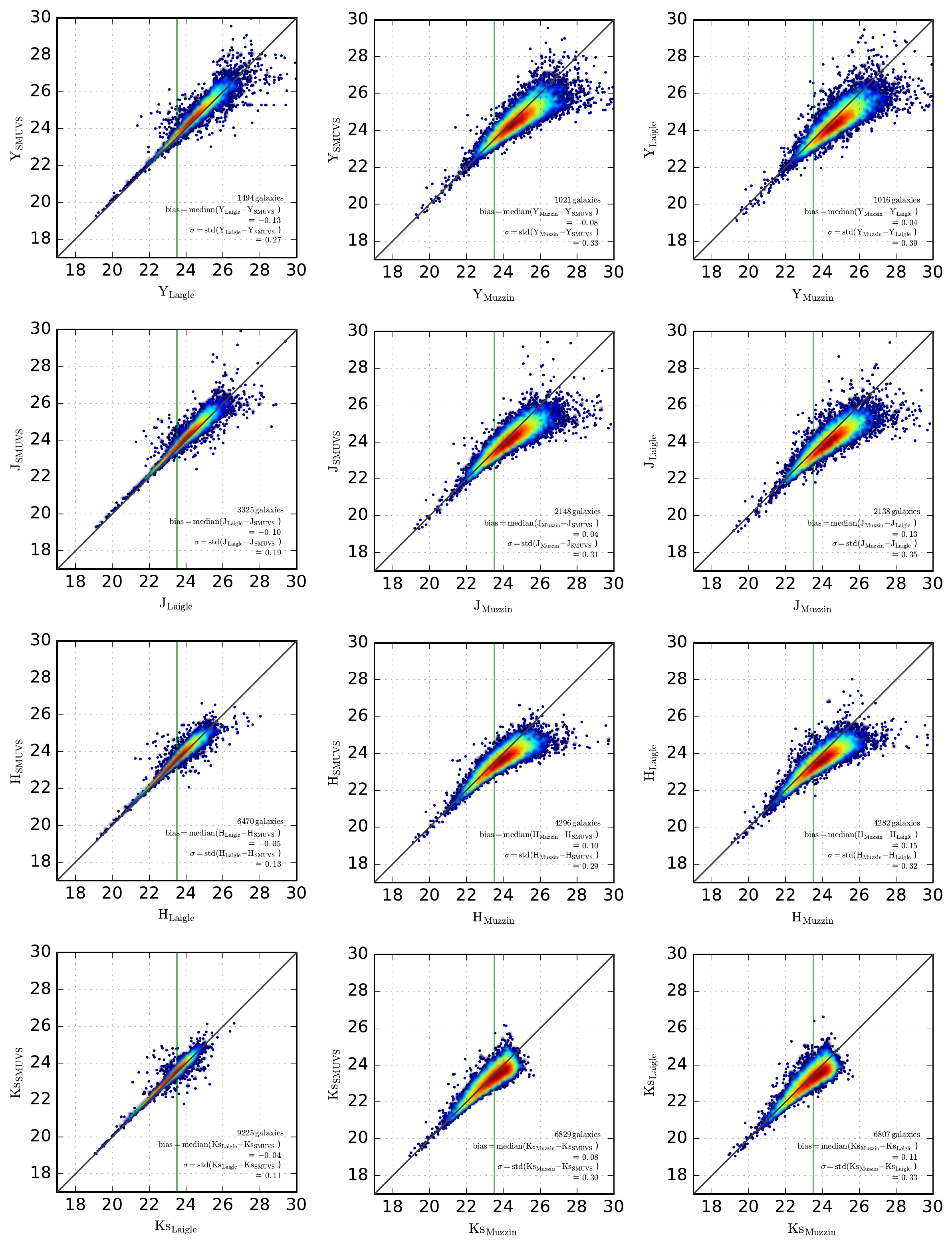}
\caption{Comparison our UltraVISTA photometry for our SMUVS $z\geq2$ galaxies with that independently obtained by \citet{muz13} and \citet{lai16} (left and middle columns), and the photometry of these two groups compared among themselves (right column). The statistics indicated in each panel label refers to sources with magnitudes $<23.5$, at which all three catalogues have a $\gsim 90\%$ completeness level.}
\label{fig:photcomp}
\end{figure*}

Our photometry on the ground-based images has been performed without a prior PSF-matching. However, as we deal here only with $z\geq2$ galaxies, we derive aperture corrections for all our magnitudes on a filter-by-filter basis, and on each stripe separately, this procedure presents no concern for our photometric measurements. To demonstrate this, in Fig.~\ref{fig:photcomp} we compare our UltraVISTA photometry for our $z>2$ SMUVS sources, with those obtained by \citet{muz13} and \citet{lai16}, who have independently performed their source photometry after PSF-matching the different ground-based images.
	
As can be seen from the different panels in Fig.~\ref{fig:photcomp}, there is an overall good agreement between the photometric measurements performed by different groups. The photometric biases, although non negligible, are $\leq 0.15$~mag in all cases.  The scatter between our photometry and that of \citet{lai16}  is small, but significantly larger when compared to the photometry of \citet{muz13}. Interestingly, the comparison between \citet{lai16} and \citet{muz13} photometries (both obtained after PSF-matching) yields a similarly large scatter level. This results allow us to conclude that PSF-matching has a minor impact on the photometry of $z>2$ sources, and is not the main source of the (mild) discrepancies seen between different photometric measurements on ground-based images.

\begin{figure*}
\centering
\includegraphics[width=0.8\textwidth]{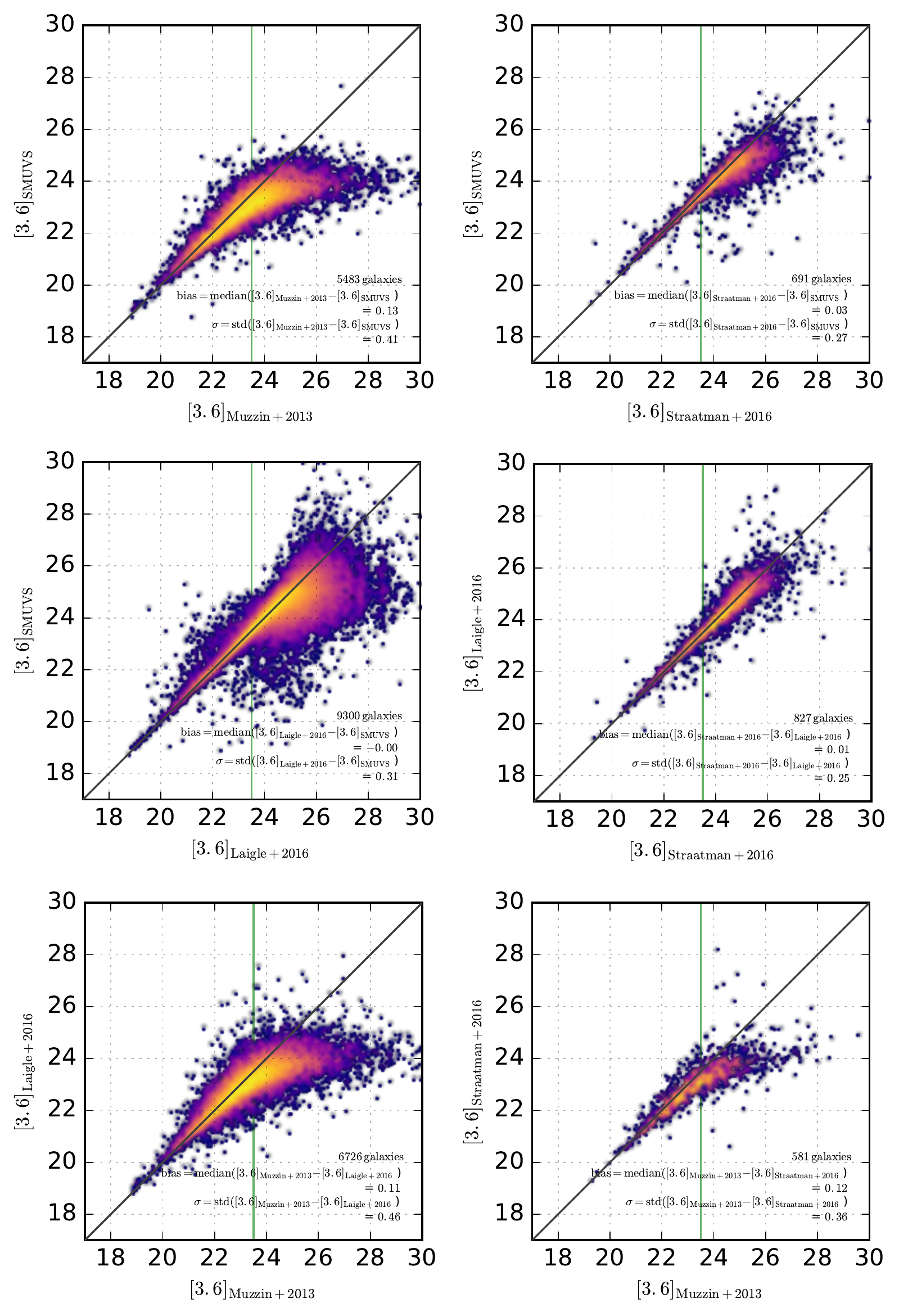}
\caption{Comparison of total magnitudes between the IRAC $3.6 \, \rm \mu m$ photometry of SMUVS $z\geq2$ sources, obtained by different authors (based on shallower COSMOS/IRAC images than the SMUVS images considered here). The statistics indicated in each panel label refers to sources with magnitudes $<23.5$.}
\label{fig:36comp}

\end{figure*}
\begin{figure*}
\centering
\includegraphics[width=0.8\textwidth]{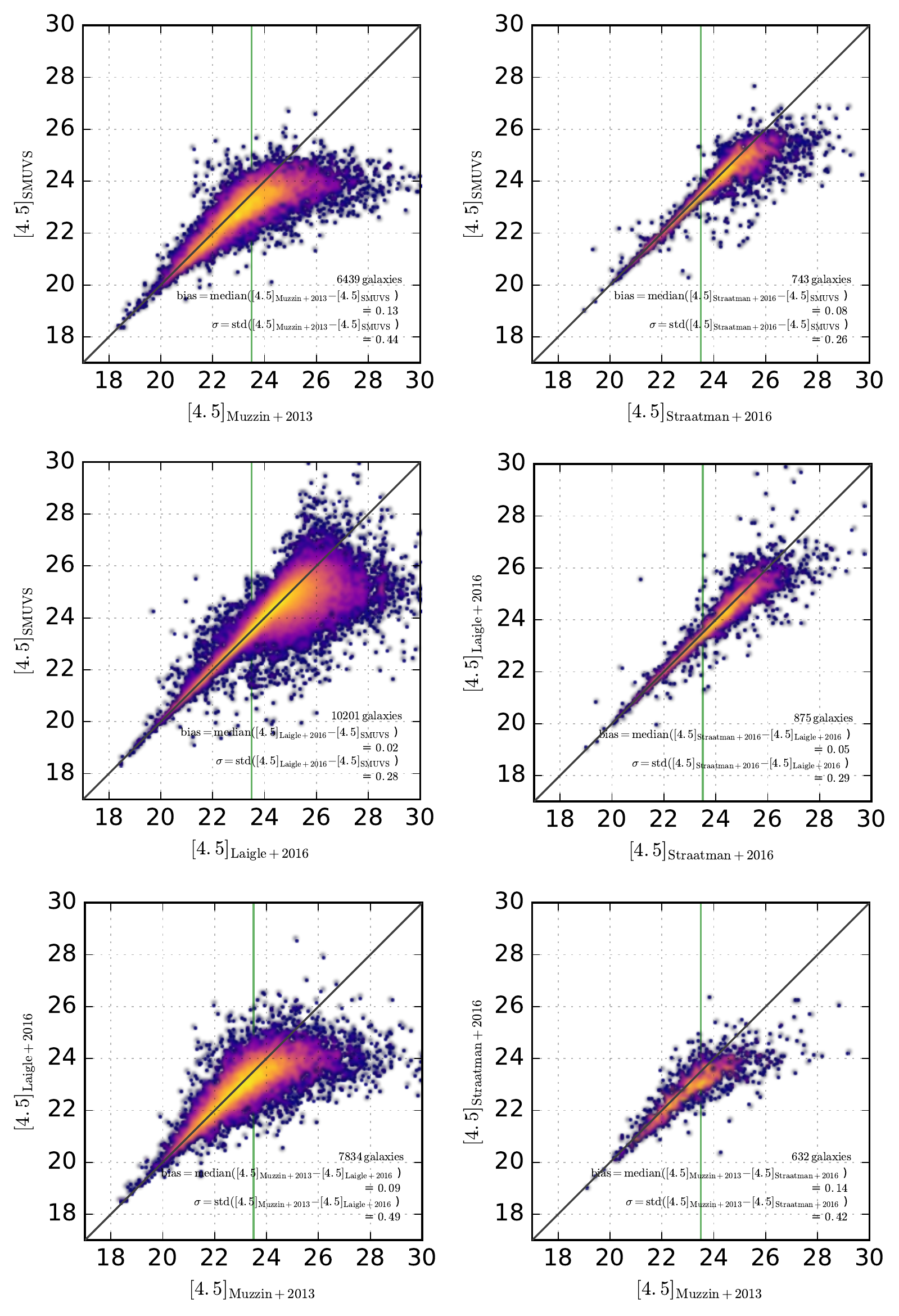}
\caption{The same as Fig.~\ref{fig:36comp}, but for IRAC $4.5 \, \rm \mu m$ photometry.}
\label{fig:45comp}
\end{figure*}

In Fig.~\ref{fig:36comp} and \ref{fig:45comp}, we show a comparison of the IRAC photometry of our sources, as measured by different groups independently \citep{muz13,lai16,2016ApJ...830...51S} on shallower COSMOS/IRAC maps, and our own from the SMUVS mosaics. As explained in \S\ref{sec:data}, our IRAC photometric measurements have been obtained with a PSF-fitting technique (using the public IRAF DAOPHOT package), which assumes that all sources are point-like. The other photometric measurements have been obtained using private codes that also fit the light profiles of all sources simultaneously, but taking into account the source shapes. In \S\ref{sec:data} we claim that taking into account source shapes is irrelevant for the vast majority of $z>2$ sources.

Indeed, this can be seen from Fig.~\ref{fig:36comp} and \ref{fig:45comp}.   Particularly, the middle left panel compares our SMUVS photometry with that in \citet{lai16} and shows a very small bias and scatter, indicating that taking into account the source shapes does not have any major impact on the IRAC photometry. Instead, other factors (e.g., recipes to re-convert fluxes from different PSF sizes, aperture corrections) may have a much more important influence on the resulting photometry.  This can be seen from the significant discrepancies among some of the photometric measurements  based on different codes that do take into account the source shapes%(with the largest discrepancies appearing in the comparisons against \citet{muz13}). 
. Surprisingly, the impact of these differences on the derived statistical galaxy properties, such as the GSMF, are very small (as can be seen in Appendix~\ref{sec:gsmfcomparison}).

\section{GSMF comparison with previous works}
\label{sec:gsmfcomparison}

\begin{figure*}
\centering
\includegraphics[width=0.8\textwidth]{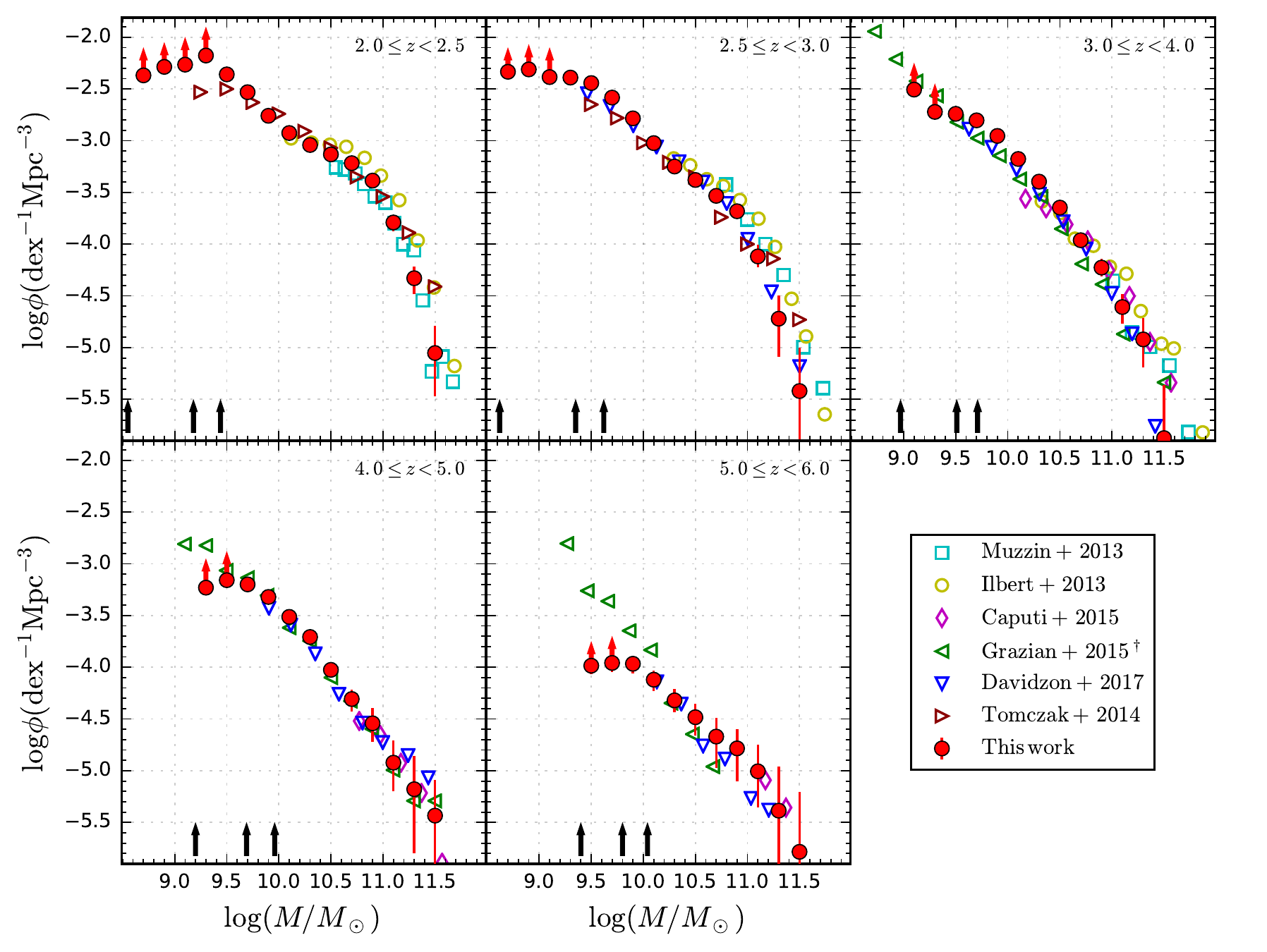}
\caption{Comparison of our total GSMF (red filled circles) with previous GSMF determinations in COSMOS and other fields. All literature GSMF were converted to a Chabrier~(2003) IMF for consistency with our own determination. $^\dagger$ Note that the data points from \citet{gra15} correspond to redshift bins with $\Delta z=+0.5$ with respect to ours. }
\label{fig:gsmflit}
\end{figure*}

As a sanity check, we computed the total GSMF at $z=2-6$ and compared our results with other recent GSMF determinations in  COSMOS and other fields.  We show our results in Fig.~\ref{fig:gsmflit}.

In the redshift range of $2.0\leq z < 2.5$, our GSMF is in generally good agreement with \cite{ilb13} and \cite{muz13} at $M_* \gsim 10^{10} \, \rm M_\odot$. No comparison is possible at lower stellar masses, as these previous works are based on shallower data. Here we are able to probe galaxies down to $\sim 1.5-2.0$~mag fainter, and we are able to clearly identify the intermediate-mass dip in the GSMF \citep[e.g.,][]{poz10}  even at these high redshifts. 

At $z=3-4$ we compare our GSMF with those determined by  \citet{cap15} and \citet{dav17} in COSMOS, and \citet{gra15} in the GOODS-South (GOODS-S) and UDS/CANDELS fields (note that the \citet{gra15} datapoints correspond to redshift bins shifted by $\Delta z = +0.5$). We see that our GSMF is in excellent agreement with these previous determinations. The SMUVS/UltraVISTA data are only $\sim 1.5$~mag shallower than the images analysed by  \citet{gra15} in the UDS and the wide GOODS-S, which results in a stellar-mass completeness limits only $\sim 0.5$~dex larger, as can be seen in Fig.~\ref{fig:gsmflit}.

At $z=4-6$, our GSMF is in very good agreement with previous works at high and intermediate stellar masses down to $\log_{10} (M_*/M_\odot) \lsim 9.2-9.5$. Below these stellar masses, our sample suffers from significant incompleteness.

\section{Errors on the GSMF}
\label{sec: gsmf unc}
\begin{figure*}
	\centering
	\includegraphics[width=0.8\textwidth]{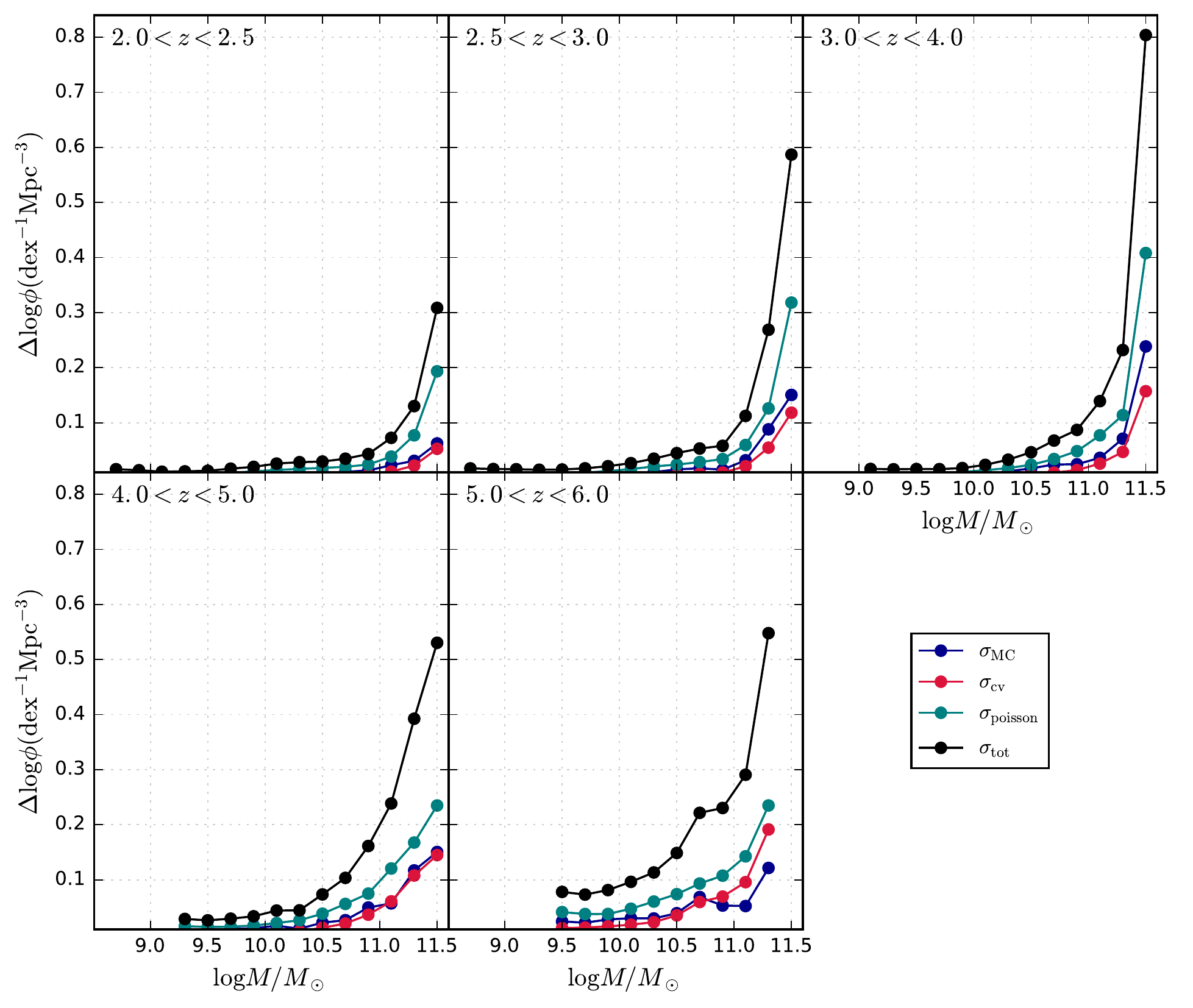}
	\caption{Contribution of each source of uncertainty to the total GSMF (Poisson ($\sigma_\mathrm{poisson}$), cosmic variance ($\sigma_\mathrm{CV}$) and uncertainties arising from Monte Carlo ($\sigma_\mathrm{MC}$) realisations of photometry.)}
	\label{fig: error budget}
\end{figure*}

In Figure \ref{fig: error budget}, we show the contribution of each source of uncertainty to the total GSMF. The uncertainties are dominated by Poissonian errors at the high-mass end. At higher redshifts, the contributions from $\sigma_\mathrm{MC}$ and $\sigma_\mathrm{cv}$ become increasingly more important.

\section{GSMF}
\label{sec: gsmf tables}

In Tables \ref{tab: gsmf 2z3}, \ref{tab: gsmf 3z5}, \ref{tab: gsmf 5z6} we present the GSMF values at $2.0\leq z <3.0$, $3.0 \leq z < 5.0$ and $5.0 \leq z \leq 6.0$. We quote the dusty, non-dusty, non-dusty blue and non-dusty red GSMF components separately. 

\begin{table}
	\centering
\begin{tabular}{c|c|c|c|c|c|c|c|c}
	\multicolumn{1}{c|}{} &
	\multicolumn{4}{c|}{$2.0\leq z < 2.5$} &
	\multicolumn{4}{c}{$2.5\leq z < 3.0$} \\ 
	\multicolumn{1}{c}{$\log M/M_\odot$} &
	\multicolumn{1}{|c|}{$\Phi_\mathrm{d}^a$} &
	\multicolumn{1}{c|}{$\Phi_\mathrm{nd}^b$} &
	\multicolumn{1}{c|}{$\Phi_\mathrm{nd-b}^c$} &
	\multicolumn{1}{c|}{$\Phi_\mathrm{nd-r}^d$} &
	\multicolumn{1}{c|}{$\Phi_\mathrm{d}$} &
	\multicolumn{1}{c|}{$\Phi_\mathrm{nd}$} &
	\multicolumn{1}{c|}{$\Phi_\mathrm{nd-b}$} &
	\multicolumn{1}{c|}{$\Phi_\mathrm{nd-r}$}	
	\\ \hline \hline

8.7 & 4.257$_{-0.359}^{+0.386}$ & 38.584$_{-1.642}^{+1.516}$ & 36.744$_{-1.672}^{+1.735}$ & 1.839$_{-0.353}^{+0.368}$ & 9.332$_{-0.587}^{+0.559}$ & 37.065$_{-1.673}^{+2.229}$ & 36.698$_{-1.849}^{+2.180}$ & 0.367$_{-0.181}^{+0.192}$ \\
8.9 & 3.646$_{-0.395}^{+0.369}$ & 48.109$_{-1.490}^{+1.641}$ & 46.382$_{-1.515}^{+1.615}$ & 1.726$_{-0.374}^{+0.315}$ & 11.756$_{-0.561}^{+0.721}$ & 37.161$_{-1.708}^{+1.754}$ & 35.923$_{-1.673}^{+1.740}$ & 1.239$_{-0.279}^{+0.322}$ \\
9.1 & 3.507$_{-0.383}^{+0.414}$ & 50.871$_{-1.310}^{+1.408}$ & 49.483$_{-1.412}^{+1.257}$ & 1.388$_{-0.251}^{+0.223}$ & 10.420$_{-0.577}^{+0.697}$ & 30.824$_{-1.135}^{+1.407}$ & 29.725$_{-1.278}^{+1.325}$ & 1.100$_{-0.268}^{+0.310}$ \\
9.3 & 3.507$_{-0.320}^{+0.398}$ & 63.118$_{-1.947}^{+1.600}$ & 61.608$_{-1.948}^{+1.641}$ & 1.510$_{-0.249}^{+0.231}$ & 7.993$_{-0.556}^{+0.521}$ & 32.784$_{-1.157}^{+1.366}$ & 31.498$_{-1.118}^{+1.369}$ & 1.287$_{-0.223}^{+0.254}$ \\
9.5 & 4.432$_{-0.423}^{+0.383}$ & 39.364$_{-1.214}^{+1.187}$ & 38.500$_{-1.330}^{+1.113}$ & 0.865$_{-0.190}^{+0.207}$ & 5.369$_{-0.457}^{+0.475}$ & 30.725$_{-1.053}^{+1.146}$ & 29.507$_{-1.054}^{+1.075}$ & 1.218$_{-0.217}^{+0.217}$ \\
9.7 & 5.972$_{-0.555}^{+0.453}$ & 23.482$_{-0.917}^{+1.003}$ & 22.747$_{-0.979}^{+0.927}$ & 0.735$_{-0.162}^{+0.175}$ & 5.020$_{-0.479}^{+0.533}$ & 21.107$_{-0.901}^{+1.014}$ & 20.108$_{-0.920}^{+0.878}$ & 1.000$_{-0.178}^{+0.205}$ \\
9.9 & 6.348$_{-0.538}^{+0.443}$ & 11.074$_{-0.605}^{+0.647}$ & 10.175$_{-0.606}^{+0.610}$ & 0.900$_{-0.160}^{+0.200}$ & 5.381$_{-0.515}^{+0.496}$ & 11.071$_{-0.616}^{+0.663}$ & 10.538$_{-0.591}^{+0.668}$ & 0.533$_{-0.133}^{+0.154}$ \\
10.1 & 6.525$_{-0.463}^{+0.458}$ & 5.347$_{-0.393}^{+0.457}$ & 4.257$_{-0.368}^{+0.384}$ & 1.090$_{-0.184}^{+0.212}$ & 4.839$_{-0.406}^{+0.499}$ & 4.646$_{-0.434}^{+0.483}$ & 4.303$_{-0.390}^{+0.409}$ & 0.343$_{-0.092}^{+0.145}$ \\
10.3 & 5.765$_{-0.503}^{+0.491}$ & 3.332$_{-0.340}^{+0.366}$ & 2.002$_{-0.273}^{+0.262}$ & 1.330$_{-0.202}^{+0.229}$ & 3.605$_{-0.343}^{+0.385}$ & 2.031$_{-0.266}^{+0.280}$ & 1.485$_{-0.270}^{+0.225}$ & 0.546$_{-0.150}^{+0.180}$ \\
10.5 & 4.476$_{-0.436}^{+0.462}$ & 2.902$_{-0.329}^{+0.301}$ & 1.457$_{-0.221}^{+0.252}$ & 1.444$_{-0.250}^{+0.207}$ & 2.716$_{-0.352}^{+0.308}$ & 1.460$_{-0.257}^{+0.230}$ & 0.863$_{-0.184}^{+0.187}$ & 0.597$_{-0.171}^{+0.158}$ \\
10.7 & 3.573$_{-0.359}^{+0.354}$ & 2.496$_{-0.294}^{+0.324}$ & 0.874$_{-0.190}^{+0.219}$ & 1.622$_{-0.233}^{+0.247}$ & 1.752$_{-0.260}^{+0.310}$ & 1.180$_{-0.215}^{+0.260}$ & 0.533$_{-0.139}^{+0.154}$ & 0.647$_{-0.184}^{+0.165}$ \\
10.9 & 2.243$_{-0.290}^{+0.321}$ & 1.875$_{-0.280}^{+0.265}$ & 0.519$_{-0.137}^{+0.167}$ & 1.356$_{-0.227}^{+0.239}$ & 1.130$_{-0.233}^{+0.234}$ & 0.952$_{-0.206}^{+0.217}$ & 0.241$_{-0.111}^{+0.130}$ & 0.711$_{-0.158}^{+0.180}$ \\
11.1 & 0.760$_{-0.183}^{+0.214}$ & 0.862$_{-0.171}^{+0.229}$ & 0.228$_{-0.088}^{+0.117}$ & 0.634$_{-0.166}^{+0.167}$ & 0.508$_{-0.140}^{+0.203}$ & 0.254$_{-0.116}^{+0.133}$ & 0.038$_{-0.038}^{+0.069}$ & 0.216$_{-0.079}^{+0.112}$ \\
11.3 & 0.203$_{-0.091}^{+0.109}$ & 0.266$_{-0.115}^{+0.117}$ & 0.025$_{-0.025}^{+0.056}$ & 0.241$_{-0.097}^{+0.113}$ & 0.152$_{-0.086}^{+0.121}$ & 0.038$_{-0.038}^{+0.062}$ & 0.013$_{-0.013}^{+0.037}$ & 0.025$_{-0.025}^{+0.058}$ \\
11.5 & 0.051$_{-0.045}^{+0.079}$ & 0.038$_{-0.038}^{+0.060}$ & 0.025$_{-0.035}^{+0.040}$ & 0.013$_{-0.013}^{+0.050}$ & 0.025$_{-0.025}^{+0.058}$ & 0.013$_{-0.013}^{+0.037}$ & - & 0.013$_{-0.013}^{+0.053}$ 

	\end{tabular}
\caption{Tabulated values of GSMFs for $2.0\leq z < 3.0$. All number densities are in units of $10^{-4} \times \mathrm{Mpc^{-3}dex^{-1} }$. Uncertainties include Poisson noise, SED-modelling uncertainties and cosmic variance.  \hspace{\textwidth} Notes: $^a$ dusty; $^b$ non-dusty all; $^c$ non-dusty blue; $^d$ non-dusty red. }
\label{tab: gsmf 2z3}
\end{table}

\begin{table}
	\centering
\begin{tabular}{c|c|c|c|c|c|c|c|c}
	\multicolumn{1}{c|}{} &
	\multicolumn{4}{c|}{$3.0\leq z < 4.0$} &
	\multicolumn{4}{c}{$4.0\leq z < 5.0$} \\ 
	\multicolumn{1}{c}{$\log M/M_\odot$} &
	\multicolumn{1}{|c|}{$\Phi_\mathrm{d}$} &
	\multicolumn{1}{c|}{$\Phi_\mathrm{nd}$} &
	\multicolumn{1}{c|}{$\Phi_\mathrm{nd-b}$} &
	\multicolumn{1}{c|}{$\Phi_\mathrm{nd-r}$} &
	\multicolumn{1}{c|}{$\Phi_\mathrm{d}$} &
	\multicolumn{1}{c|}{$\Phi_\mathrm{nd}$} &
	\multicolumn{1}{c|}{$\Phi_\mathrm{nd-b}$} &
	\multicolumn{1}{c|}{$\Phi_\mathrm{nd-r}$}	
	\\ \hline \hline

9.1 & 4.867$_{-0.325}^{+0.295}$ & 26.366$_{-1.264}^{+1.368}$ & 25.553$_{-1.085}^{+1.143}$ & 0.813$_{-0.173}^{+0.213}$ & - & - & - & - \\
9.3 & 3.990$_{-0.310}^{+0.271}$ & 15.005$_{-0.587}^{+0.567}$ & 13.840$_{-0.587}^{+0.519}$ & 1.165$_{-0.161}^{+0.185}$ & 2.665$_{-0.256}^{+0.263}$ & 3.223$_{-0.308}^{+0.318}$ & 3.104$_{-0.318}^{+0.277}$ & 0.119$_{-0.075}^{+0.108}$ \\
9.5 & 2.786$_{-0.285}^{+0.260}$ & 15.416$_{-0.567}^{+0.546}$ & 14.027$_{-0.500}^{+0.541}$ & 1.389$_{-0.194}^{+0.180}$ & 2.689$_{-0.226}^{+0.236}$ & 4.266$_{-0.336}^{+0.317}$ & 3.807$_{-0.322}^{+0.307}$ & 0.458$_{-0.96}^{+0.125}$ \\
9.7 & 2.137$_{-0.220}^{+0.228}$ & 13.609$_{-0.503}^{+0.583}$ & 12.329$_{-0.486}^{+0.566}$ & 1.280$_{-0.177}^{+0.153}$ & 1.878$_{-0.259}^{+0.211}$ & 4.451$_{-0.317}^{+0.351}$ & 3.902$_{-0.261}^{+0.339}$ & 0.549$_{-0.119}^{+0.145}$ \\
9.9 & 2.137$_{-0.240}^{+0.223}$ & 8.991$_{-0.405}^{+0.405}$ & 8.265$_{-0.406}^{+0.375}$ & 0.726$_{-0.123}^{+0.114}$ & 1.435$_{-0.933}^{+0.228}$ & 3.350$_{-0.305}^{+0.302}$ & 2.690$_{-0.279}^{+0.242}$ & 0.659$_{-0.128}^{+0.131}$ \\
10.1 & 1.897$_{-0.253}^{+0.191}$ & 4.764$_{-0.333}^{+0.318}$ & 4.389$_{-0.321}^{+0.289}$ & 0.375$_{-0.083}^{+0.091}$ & 0.965$_{-0.181}^{+0.165}$ & 2.106$_{-0.236}^{+0.260}$ & 1.693$_{-0.239}^{+0.230}$ & 0.413$_{-0.093}^{+0.104}$ \\
10.3 & 1.735$_{-0.193}^{+0.205}$ & 2.290$_{-0.215}^{+0.217}$ & 2.057$_{-0.207}^{+0.203}$ & 0.234$_{-0.066}^{+0.074}$ & 0.692$_{-0.133}^{+0.141}$ & 1.271$_{-0.171}^{+0.169}$ & 1.018$_{-0.145}^{+0.173}$ & 0.252$_{-0.068}^{+0.086}$ \\
10.5 & 1.139$_{-0.160}^{+0.186}$ & 1.118$_{-0.146}^{+0.152}$ & 0.924$_{-0.136}^{+0.129}$ & 0.194$_{-0.063}^{+0.072}$ & 0.354$_{-0.115}^{+0.116}$ & 0.591$_{-0.125}^{+0.167}$ & 0.385$_{-0.091}^{+0.137}$ & 0.206$_{-0.074}^{+0.076}$ \\
10.7 & 0.694$_{-0.133}^{+0.133}$ & 0.398$_{-0.100}^{+0.108}$ & 0.238$_{-0.070}^{+0.079}$ & 0.160$_{-0.062}^{+0.071}$ & 0.339$_{-0.105}^{+0.104}$ & 0.155$_{-0.069}^{+0.081}$ & 0.118$_{-0.062}^{+0.078}$ & 0.037$_{-0.037}^{+0.049}$ \\
10.9 & 0.247$_{-0.092}^{+0.085}$ & 0.347$_{-0.089}^{+0.090}$ & 0.207$_{-0.068}^{+0.077}$ & 0.140$_{-0.052}^{+0.069}$ & 0.214$_{-0.084}^{+0.095}$ & 0.074$_{-0.049}^{+0.060}$ & 0.044$_{-0.032}^{+0.053}$ & 0.029$_{-0.029}^{+0.040}$ \\
11.1 & 0.140$_{-0.055}^{+0.079}$ & 0.107$_{-0.057}^{+0.058}$ & 0.053$_{-0.035}^{+0.045}$ & 0.053$_{-0.039}^{+0.049}$ & 0.098$_{-0.055}^{+0.069}$ & 0.022$_{-0.022}^{+0.039}$ & 0.015$_{-0.015}^{+0.036}$ & 0.007$_{-0.007}^{+0.023}$ \\
11.3 & 0.100$_{-0.057}^{+0.058}$ & 0.020$_{-0.020}^{+0.033}$ & 0.020$_{-0.020}^{+0.033}$ & - & 0.066$_{-0.052}^{+0.062}$ & - & - & - \\
11.5 & 0.007$_{-0.007}^{+0.028}$ & 0.007$_{-0.007}^{+0.028}$ & - & 0.007$_{-0.007}^{+0.028}$ & 0.037$_{-0.037}^{+0.045}$ & - & - & - 	
	
\end{tabular}
\caption{Tabulated values of GSMFs for $3.0\leq z < 5.0$.}
\label{tab: gsmf 3z5}	
\end{table}

\begin{table}
	\centering
\begin{tabular}{c|c|c|c|c}
	\multicolumn{1}{c|}{} &
	\multicolumn{4}{c}{$5.0\leq z \leq 6.0$} \\ 
	\multicolumn{1}{c}{$\log M/M_\odot$} &
	\multicolumn{1}{|c|}{$\Phi_\mathrm{d}$} &
	\multicolumn{1}{c|}{$\Phi_\mathrm{nd}$} &
	\multicolumn{1}{c|}{$\Phi_\mathrm{nd-b}$} &
	\multicolumn{1}{c}{$\Phi_\mathrm{nd-r}$}	
	\\ \hline \hline
	
9.5 & 0.197$_{-0.079}^{+0.097}$ & 0.839$_{-0.182}^{+0.207}$ & 0.810$_{-0.165}^{+0.206}$ & 0.029$_{-0.032}^{+0.043}$ \\
9.7 & 0.230$_{-0.084}^{+0.094}$ & 0.871$_{-0.167}^{+0.179}$ & 0.808$_{-0.173}^{+0.176}$ & 0.064$_{-0.048}^{+0.052}$ \\
9.9 & 0.206$_{-0.071}^{+0.104}$ & 0.876$_{-0.163}^{+0.189}$ & 0.773$_{-0.140}^{+0.183}$ & 0.103$_{-0.056}^{+0.070}$ \\
10.1 & 0.115$_{-0.071}^{+0.082}$ & 0.644$_{-0.171}^{+0.161}$ & 0.586$_{-0.151}^{+0.141}$ & 0.058$_{-0.047}^{+0.049}$ \\
10.3 & 0.107$_{-0.059}^{+0.069}$ & 0.372$_{-0.100}^{+0.129}$ & 0.282$_{-0.091}^{+0.126}$ & 0.091$_{-0.054}^{+0.058}$ \\
10.5 & 0.058$_{-0.058}^{+0.062}$ & 0.272$_{-0.110}^{+0.111}$ & 0.222$_{-0.056}^{+0.100}$ & 0.049$_{-0.047}^{+0.060}$ \\
10.7 & 0.066$_{-0.054}^{+0.067}$ & 0.148$_{-0.079}^{+0.084}$ & 0.115$_{-0.064}^{+0.077}$ & 0.033$_{-0.033}^{+0.047}$ \\
10.9 & 0.115$_{-0.067}^{+0.080}$ & 0.049$_{-0.040}^{+0.066}$ & 0.049$_{-0.040}^{+0.054}$ & - \\
11.1 & 0.074$_{-0.061}^{+0.063}$ & 0.025$_{-0.025}^{+0.036}$ & 0.025$_{-0.032}^{+0.036}$ & - \\
11.3 & 0.033$_{-0.033}^{+0.066}$ & 0.008$_{-0.008}^{+0.042}$ & 0.008$_{-0.008}^{+0.042}$ & - \\
11.5 & - & 0.016$_{-0.016}^{+0.046}$ & 0.008$_{-0.008}^{+0.042}$ & 0.008$_{-0.008}^{+0.029}$ 

\end{tabular}	
\caption{Tabulated values of GSMFs for $5.0\leq z < 6.0$.}
\label{tab: gsmf 5z6}	
\end{table}

\bibliography{sdeshmukh_smuvscatgsmf.bib}

%\end{thebibliography}

\end{document}